\documentclass{article} 
\usepackage{iclr2023_conference,times}

\usepackage{amsmath,amsfonts,bm}

\def\eqref#1{equation~\ref{#1}}

\def\floor#1{\lfloor #1 \rfloor}
\def\1{\bm{1}}

\def\rs{{\textnormal{s}}}

\DeclareMathAlphabet{\mathsfit}{\encodingdefault}{\sfdefault}{m}{sl}
\SetMathAlphabet{\mathsfit}{bold}{\encodingdefault}{\sfdefault}{bx}{n}

\newcommand{\E}{\mathbb{E}}

\usepackage{hyperref}
\usepackage{url}
\usepackage{amsmath}  
\usepackage{rotating}  
\usepackage{adjustbox}  
\usepackage{pdflscape}  
\usepackage{textcomp}  
\usepackage{amsfonts}  
\usepackage{bm}  
\usepackage{amsthm}  
\usepackage{mathrsfs}  
\usepackage{thmtools, thm-restate}  
\usepackage{graphicx}  
\usepackage{subfigure}  
\usepackage{chngcntr}  
\usepackage{makecell}  
\usepackage{wrapfig}  
\usepackage[usestackEOL]{stackengine}  
\usepackage{amssymb}   
\usepackage{readarray}  
\usepackage{xspace}  
\usepackage{float}  
\usepackage{multirow}  
\usepackage{booktabs} 
\usepackage[algo2e,linesnumbered,noend,ruled]{algorithm2e}
\usepackage{pifont}   
\usepackage{caption,array}  
\usepackage[T1]{fontenc}  
\usepackage{mathtools}

\usepackage{xstring}  
\usepackage{microtype}
\usepackage{breakurl}
\usepackage{xcolor}
\usepackage{footnote}
\usepackage{tikz}
\usepackage{collcell}  
\usepackage{mleftright}
\usepackage{etoolbox}
\usepackage{tikz}
\usepackage[normalem]{ulem}
\usepackage[capitalise,noabbrev]{cleveref}

\usepackage{pgfplots}
\pgfkeys{/pgf/fpu}
\pgfmathparse{16383+1}

\pgfkeys{/pgf/fpu=false}

\usepackage{xr}
\makeatletter
\newcommand*{\addFileDependency}[1]{
  \typeout{(#1)}
  \@addtofilelist{#1}
  \IfFileExists{#1}{}{\typeout{No file #1.}}
}
\makeatother

\newcommand*{\myexternaldocument}[1]{
    \externaldocument{#1}
    \addFileDependency{#1.tex}
    \addFileDependency{#1.aux}
}
\myexternaldocument{appendix}

\newcommand{\ie}{\textit{i.e.}}
\newcommand{\eg}{\textit{e.g.}}

\newcommand{\rpmnotation}{\Psi}  

\newcommand{\cmark}{\textcolor{green}{\ding{51}}}%
\newcommand{\xmark}{\textcolor{gray}{\ding{55}}}%

\theoremstyle{plain}

\theoremstyle{definition}

\DeclareMathAlphabet\mathbfcal{OMS}{cmsy}{b}{n}

\definecolor{goodcolor}{RGB}{200,200,255}
\definecolor{midcolor}{RGB}{240,240,240}
\definecolor{badcolor}{RGB}{255,200,200}

\newcounter{MinX}
\newcounter{MidX}
\newcounter{MaxX}
\newcommand{\ApplyPositiveGradient}[1]{%
    \IfDecimal{#1}{%
        \pgfmathsetmacro{\Input}{#1 * 100}%
        \ifdim \Input pt > \value{MidX} pt%
            \pgfmathsetmacro{\PercentColor}{max(min(pow((\Input - \value{MidX})/(\value{MaxX}-\value{MidX}),0.75),1.0),0.00)*100}%
            \colorbox{goodcolor!\PercentColor!midcolor}{#1}%
        \else%
            \pgfmathsetmacro{\PercentColor}{max(min(pow((\value{MidX} - \Input)/(\value{MidX}-\value{MinX}),0.75),1.0),0.00)*100}%
            \colorbox{badcolor!\PercentColor!midcolor}{#1}%
        \fi%
    }{\bf #1}%
}

\newcommand{\ApplyGradient}[1]{%
    \IfDecimal{#1}{%
        \pgfmathsetmacro{\Input}{#1}%
        \ifdim \Input pt > \value{MidX} pt%
            \pgfmathsetmacro{\PercentColor}{max(min(pow((\Input - \value{MidX})/(\value{MaxX}-\value{MidX}),0.75),1.0),0.00)*100}%
            \colorbox{goodcolor!\PercentColor!midcolor}{#1}%
        \else%
            \pgfmathsetmacro{\PercentColor}{max(min(pow((\value{MidX} - \Input)/(\value{MidX}-\value{MinX}),0.75),1.0),0.00)*100}%
            \colorbox{badcolor!\PercentColor!midcolor}{#1}%
        \fi%
    }{\bf #1}%
}

\newcommand{\ApplyNegativeGradient}[1]{%
    \IfDecimal{#1}{%
        \pgfmathsetmacro{\Input}{#1 * 100}%
        \ifdim \Input pt > \value{MidX} pt%
            \pgfmathsetmacro{\PercentColor}{max(min(100.0*pow((\Input - \value{MidX})/(\value{MaxX}-\value{MidX}),0.75),100.0),0.00)}
            \colorbox{badcolor!\PercentColor!midcolor}{#1}%
        \else%
            \pgfmathsetmacro{\PercentColor}{max(min(100.0*pow((\value{MidX} - \Input)/(\value{MidX}-\value{MinX}),0.75),100.0),0.00)}%
            \colorbox{goodcolor!\PercentColor!midcolor}{#1}%
        \fi%
    }{\bf #1}%
}

\SetAlFnt{\footnotesize}
\SetAlCapFnt{\footnotesize}
\SetAlCapNameFnt{\footnotesize}
\usepackage{bbm}   
\usepackage[most]{tcolorbox}
\tcbset{
    frame code={}
    center title,
    left skip=0pt,
    left=0pt,
    leftrule=-2pt,
    rightrule=-2pt,
    right=0pt,
    top=0pt,
    bottom=0pt,
    colback=black!4,
    width=1\textwidth,
    boxsep=5pt,
    arc=0pt,outer arc=0pt,
    }
\newcommand{\thmbox}[2]{\begin{tcolorbox}\begin{#1} #2 \end{#1}\end{tcolorbox}}
\newcommand*{\rom}[1]{\expandafter\@slowromancap\romannumeral #1@}

\usepackage{ifthen}
\newboolean{include-notes}
\setboolean{include-notes}{true}
\newcommand{\nb}[3]{\ifthenelse{\boolean{include-notes}}{{\colorbox{#2}{\bfseries\sffamily\scriptsize\textcolor{white}{#1}}}{\ \textcolor{#2}{\sf\small\textit{#3}}}}{}}

\title{RPM: Generalizable Behaviors for Multi-Agent Reinforcement Learning}

\iclrfinalcopy 

\author{Wei Qiu\thanks{This work was done during an internship at Sea AI Lab, Singapore. \\Correspondence to: Wei Qiu <\href{mailto:qiuw0008@e.ntu.edu.sg}{qiuw0008@e.ntu.edu.sg}> and Zhongwen Xu <\href{mailto:xuzw@sea.com}{xuzw@sea.com}>.}~~\textsuperscript{\textnormal{1}~\textnormal{2}}, Xiao Ma~\textsuperscript{\textnormal{2}}, Bo An~\textsuperscript{\textnormal{1}}, Svetlana Obraztsova~\textsuperscript{\textnormal{1}}, Shuicheng Yan~\textsuperscript{\textnormal{2}}, Zhongwen Xu~\textsuperscript{\textnormal{2}} \\
\textsuperscript{1}Nanyang Technological University, Singapore \\
\textsuperscript{2}Sea AI Lab, Singapore\\
}

\newcommand{\incres}{402}
\begin{document}

\maketitle

\begin{abstract}
Despite the recent advancement in multi-agent reinforcement learning (MARL), the MARL agents easily overfit the training environment and perform poorly in the evaluation scenarios where other agents behave differently. 
Obtaining generalizable policies for MARL agents is thus necessary but challenging mainly due to complex multi-agent interactions. 
In this work, we model the problem with Markov Games and propose a simple yet effective method, ranked policy memory (RPM), to collect diverse multi-agent trajectories for training MARL policies with good generalizability.
The main idea of RPM is to maintain a look-up memory of policies. 
In particular, we try to acquire various levels of behaviors by saving policies via ranking the training episode return, \ie, the episode return of agents in the training environment; when an episode starts, the learning agent can then choose a policy from the RPM as the behavior policy.
This innovative self-play training framework leverages agents' past policies and guarantees the diversity of multi-agent interaction in the training data.
We implement RPM on top of MARL algorithms and conduct extensive experiments on Melting Pot. 
It has been demonstrated that RPM enables MARL agents to interact with unseen agents in multi-agent generalization evaluation scenarios and complete given tasks, and it significantly boosts the performance up to \incres\% on average.
\end{abstract}

\section{Introduction}\label{Intro}
Recent years have witnessed considerable progress in Multi-Agent Reinforcement Learning (MARL) research~\citep{yang2020overview}. In MARL, each agent acts decentrally and interacts with other agents to complete particular tasks or achieve specific goals via reinforcement learning (RL). However, generalization~\citep{hupkes2020compositionality} remains a critical issue in MARL research. Generalization may have different meanings from different perspectives and research directions. Such as generalization at the representation level studied in supervised learning and generalization to different test environments rather than the training environments for a single agent~\citep{kirk2021survey}. 
In this work, we study the the generalization ability of agents to collaborate or compete with other agents with unseen policies during training. Such a setup is critical to real-world MARL applications~\citep{leibo2021scalable}.
Unfortunately, current MARL methods mostly neglect generalization issues and could be fragile; for example, a co-player changing its policy may cause the trained agent to fail to cooperate. 

\begin{figure}
    \vspace{-1.3cm}
    \centering
    \includegraphics[scale=0.45]{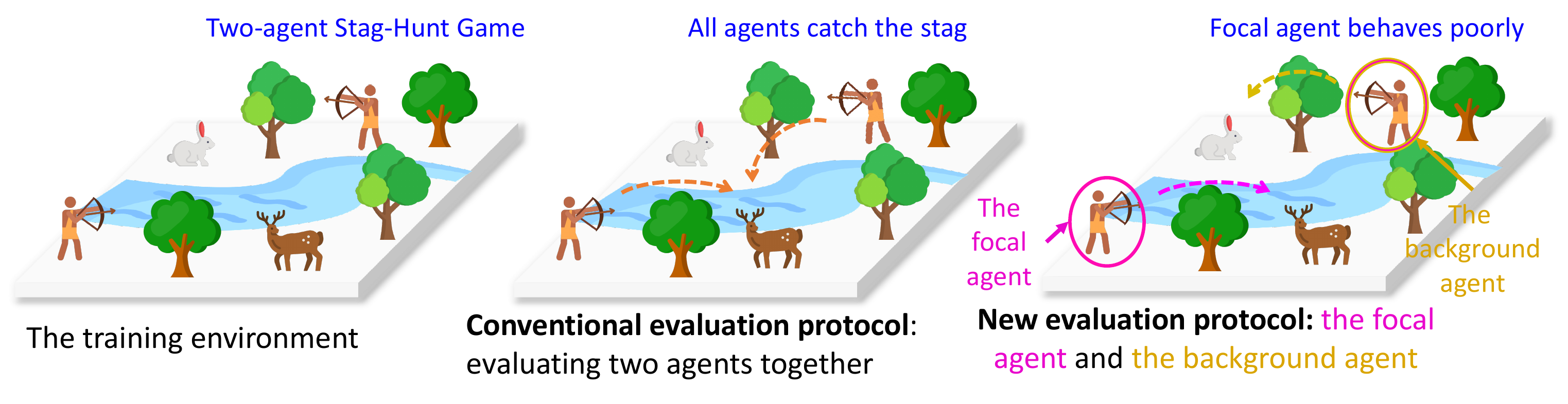}
    \vspace{-0.8cm}
    \caption{\footnotesize{Two-Agent Stag-Hunt Game. \textbf{Left:} Training environment. Two agents (hunters) hunting in the environment.
    \textbf{Middle:} After training in the training environment, all agents behave cooperatively to capture the stag for maximizing the group reward. \textbf{Right:} In the new evaluation scenario, one agent is picked as the focal agent (in the magenta circle) and paired with a pre-trained agent (in brown circle) that behaves in different ways to evaluate the performance of the selected agent. The conventional evaluation protocol fails to evaluate such behavior and current MARL easily fails to learn the optimal policy.}}
    \vspace{-0.4cm}
    \label{fig:social_example}
\end{figure}

In this work, we aim to train MARL agents that can adapt to new scenarios where other agents' policies are unseen during training for MARL. To understand the difficulties and why it is crucial to tackle generalization, we illustrate a two-agent stag-hunt game as an example in Figure~\ref{fig:social_example}. The agents are trained to obtain policies maximizing the group reward by shooting arrows to the stag. As a result, they may perform well in evaluation scenarios similar to the training environment, as shown in Figure~\ref{fig:social_example} (left) and (middle), respectively. 
However, these agents may fail when evaluated in scenarios different from the training scenarios.
As shown in Figure~\ref{fig:social_example} (right), the learning agent (called the focal agent following the convention in \citep{leibo2021scalable}) is supposed to work together with another agent (called the background agent following the naming in \citep{leibo2021scalable}) are pre-trained to be selfish (\ie, only capture the hare). In this case, the focal agent will fail to capture the stag without the help from its teammates and the optimal policy to capture the hare. However, background agents are unseen to the focal agent during training. Therefore, without generalization, the agents trained as Figure~\ref{fig:social_example} (left) cannot achieve an optimal policy in the new evaluation scenario.

We model the problem with Markov games~\citep{littman1994markov} and propose a simple yet effective method called ranked policy memory (RPM) to attain generalizable policies in multi-agent systems during training. The core idea of RPM is to maintain a look-up memory of policies during training for the agents. In particular, we first evaluate the trained agents' policies after each training update. We then rank and save the trained agents' policies by the training episode returns. In this way, we obtain various levels, \ie, performances, of policies.
When starting an episode, the agent can access the memory and load the randomly sampled policy to replace the current behavior policy. 
The new ensemble of policies enables agents in the self-play framework to collect diversified experiences in the training environment for training. These diversified experiences contain many novel multi-agent interactions that enhance the extrapolation capacity of MARL, boosting the generalization performance. 
We note that an easy extension to having different behavior properties as the keys in RPM could potentially further enrich the generalization but it is left for future work.

We implement RPM on top of state-of-the-art MARL algorithm, MAPPO~\citep{yu2021surprising}. To verify its effectiveness, we use Melting Pot~\citep{leibo2021scalable} as our testbeds. We then conduct large-scale experiments with the Melting Pot benchmark, which is a well-recognized benchmark for MARL generalization evaluation.
The experiment results demonstrate that RPM significantly boosts the performance of generalized social behaviors up to \incres\% on average and outperforms many baselines in a variety of multi-agent generalization evaluation scenarios. 
Our code, pictorial examples, and videos are available at this link: \url{https://sites.google.com/view/rpm-2022/}.

\section{Preliminaries}\label{sec:background}
\subsection{Markov Games and Notations}

We consider the Markov Games~\citep{littman1994markov} represented by a tuple $\mathcal{G}=\langle\mathcal{N}, \mathcal{S},\mathcal{A},\mathcal{O}, P, R,\gamma, \rho\rangle$. 
$\mathcal{N}$ is a set of agents with the size  $|\mathcal{N}|=N$; $\mathcal{S}$ is a set of states; $\mathcal{A}=\times_{i=1}^{N}\mathcal{A}_{i}$ is a set of joint actions with $\mathcal{A}_{i}$ denoting the set of actions for an agent $i$; 
$\mathcal{O}=\times_{i=1}^{N}\mathcal{O}_{i}$ is the observation set, with $\mathcal{O}_{i}$ denoting the observation set of the agent $i$; $P:\mathcal{S}\times\mathcal{A}\rightarrow\mathcal{S}$ is the transition function and $R=\times_{i=1}^{N}r_{i}$ is the reward function where $r_{i}:\mathcal{S}\times\mathcal{A}\rightarrow\mathbb{R}$ specifies the reward for the agent $i$ given the state and the joint action; $\gamma$ is the discount factor; the initial states are determined by a distribution $\rho:\mathcal{S}\rightarrow [0,1]$.  
Given a state $s\in\mathcal{S}$, each agent $i\in\mathcal{N}$ chooses its action $u_{i}$ and obtains the reward $r(s,\bm{u})$ with the private observation $o_{i}\in\mathcal{O}_{i}$, where $\bm{u}=\{ u_{i}\}^{N}_{i=1}$ is the joint action.
The joint policy of agents is denoted as $\bm{\pi}_{\bm{\theta}}=\{\pi_{\theta_{i}}\}^{N}_{i=1}$ where $\pi_{\theta_{i}}:\mathcal{S}\times\mathcal{A}_{i}\rightarrow[0,1]$ is the policy for the agent $i$. The objective of each agent is to maximize its own total expected return $R_{i}=\sum_{t=0}^{\infty}\gamma^{t}r^{t}_{i}$.

\subsection{Multi-Agent Reinforcement Learning}

In MARL, multiple agents act in the environment to maximize their respective returns with RL~\citep{sutton2018reinforcement}.
Each agent's policy $\pi_i$ is optimized by maximizing the following objective:
\begin{align}\nonumber
    \mathcal{J}({\pi_{i}}) \triangleq \E_{\rs_{0:\infty}\sim \rho^{0:\infty}_{\mathcal{G}}, a^{i}_{0:\infty}\sim\pi_{i}}\left[ \sum_{t=0}^{\infty}\gamma^{t}r^{i}_{t} \right],
\end{align}
where $\mathcal{J}({\pi_{i}})$ is a performance measure for policy gradient RL methods~\citep{williams1992simple,lillicrap2015continuous,fujimoto2018addressing}. Each policy's Q value
$Q_i$ is optimized by minimizing the following regression loss~\citep{watkins1992q,mnih2015human} with TD-learning:
\begin{align}\nonumber
   \mathcal{L} (\theta_i) \triangleq \mathbb{E}_{\mathcal{D}^{\prime} \sim \mathcal{D}}\left[\left(y_{t}^{i}-Q_{\theta_i}^{i}\left(\boldsymbol{s}_{t}, \boldsymbol{u}_{t}, s^{i}_{t}, u^{i}_{t}\right)\right)^{2}\right],
\end{align}
where $y_{t}^{i}=r^{i}_{t}+\gamma \max _{\boldsymbol{u}^{\prime}} Q_{\bar{\theta_i}}^{i}\left(\boldsymbol{s}_{t+1}, \boldsymbol{u}^{\prime}, s^{i}_{t}, u^{i,\prime}\right)$. $\theta_i$ are the parameters of the agents. $\bar{\theta}_i$ is the parameter of the target $Q^{i}$ and periodically copied from $\theta$. $\mathcal{D}^{\prime}$ is a sample from the replay buffer $\mathcal{D}$.

\begin{figure}
\vspace{-0.6cm}
    \begin{center}
        \centering
        \includegraphics[width=.99\textwidth]{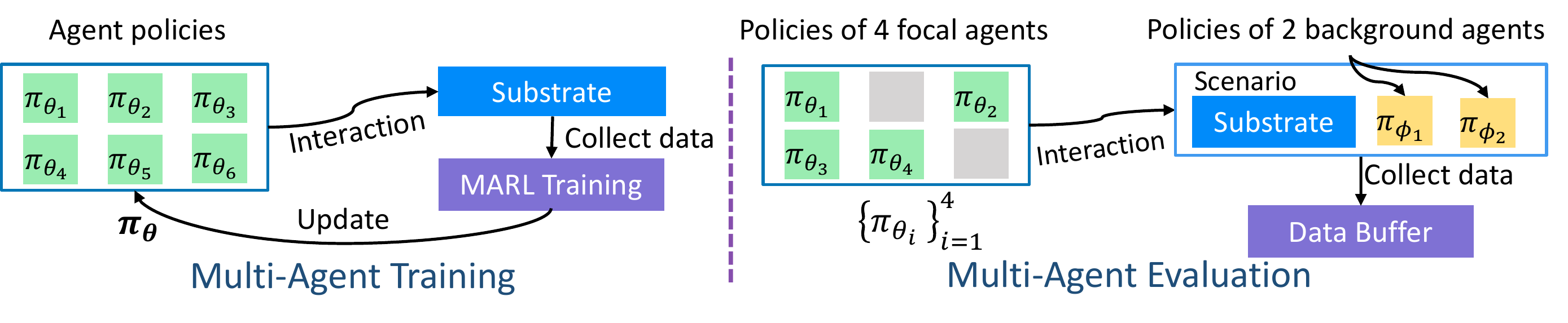}
        \vspace{-0.2cm}
        \caption{An example of our formulation. \textbf{Left:} All six agents' policies are trained with the MARL method. \textbf{Right:} Two agents with policies $\pi_{\phi_1}$ and $\pi_{\phi_2}$ are selected as background agents and the rest of the 4 agents (with new indices) are focal agents to be evaluated. The substrate and the background agents constitute the evaluation scenario }\label{fig:formulation_example}
    \vspace{-0.2cm}
    \end{center}
\end{figure}

\section{Problem Formulation}\label{sec:formulation}
We introduce the formulation of MARL for training and evaluation in our problem.
Our goal is to improve generalizabiliby of MARL policies in scenarios where policies of agents or opponents are unseen during training while the physical environment is unchanged. Following~\cite{leibo2021scalable}, the training environment is defined as \textit{substrate}. Each substrate is an $N$-agent partially observable Markov game $\mathcal{G}$. Each agent optimizes its policy $\pi_{\theta_i}$ via the following protocol.

\thmbox{definition}{[Multi-Agent Training] There are $N$ agents act in the substrate, which is denoted as $\mathcal{G}$. Each agent receives partial environmental observation not known to other agents and aims to optimizes its policy $\pi_{\theta_i}$ by optimizing its accumulated rewards: $\sum_{t=0}^{\infty}\gamma^{t}r^{i}_{t}$. The performance of the joint policy $\bm{\pi}_{\bm{\theta}}=\{\pi_{\theta_{i}}\}^{N}_{i=1}$ is measured by the mean individual return: $\bar{R}(\bm{\pi}_{\bm{\theta}}) = \frac{1}{N}\sum^{N}_{i=1} R(\pi_{\theta_i};\mathcal{G})$. $R(\pi_{\theta_i};\mathcal{G})$ measures the episode return of policy $\pi_{\theta_i}$ in game $\mathcal{G}$ for agent $i$.
\label{defintion:multi_agent_training}
}

In order to evaluate the trained MARL policies in evaluation scenario $\mathcal{G}^{\prime}$, we follow the evaluation protocol defined by ~\cite{leibo2021scalable}:

\thmbox{definition}{[Multi-Agent Evaluation] There are $M~(1\leq M \leq N-1)$ focal agents that are selected from $N$ agents. The focal agents are agents to be evaluated in evaluation scenarios. They are paired with $N-M$ background agents whose policies $\bm{\pi_{\phi}}=\{\pi_{\phi_{j}}\}^{N-M}_{j=1}$ were pre-trained with pseudo rewards in the same physical environment where the policies $\bm{\pi_{\theta}}$ are trained. To measure the generalized performance in evaluation scenarios, we use the mean individual return of focal agents as the performance measure: $\bar{R}(\{\pi_\theta\}^{M}_{i=1}) = \frac{1}{M}\sum^{M}_{i=1} R(\pi_{\theta_i};\mathcal{G}^{\prime})$.
\label{defintion:multi_agent_evaluation}
}

We show an example of our formulation in Figure~\ref{fig:formulation_example}. Note that the focal agents cannot utilise the interaction data collected during evaluation to train or finetune their policies. Without training the policies of focal agents with the collected trajectories during evaluation, the focal agents should behave adaptively to interact with the background agents to complete challenging multi-agent tasks. It is also worth noting that the ad-hoc team building~\citep{stone2010teach,gu2021online} is different from our formulation both in the training and evaluation. We discuss the differences in the related works section (Paragraph 3, Section~\ref{sec:relatedworks}).

\section{Methodology}\label{sec:method}

We propose a Ranked Policy Memory (RPM) method to provide diversified multi-agent behaviors for self-play to improve generalization of MARL. Then, we incorporate RPM into MAPPO~\citep{yu2021surprising} for training MARL policies.

\begin{figure}
\vspace{-0.6cm}
    \begin{center}
        \centering
        \includegraphics[width=1\textwidth]{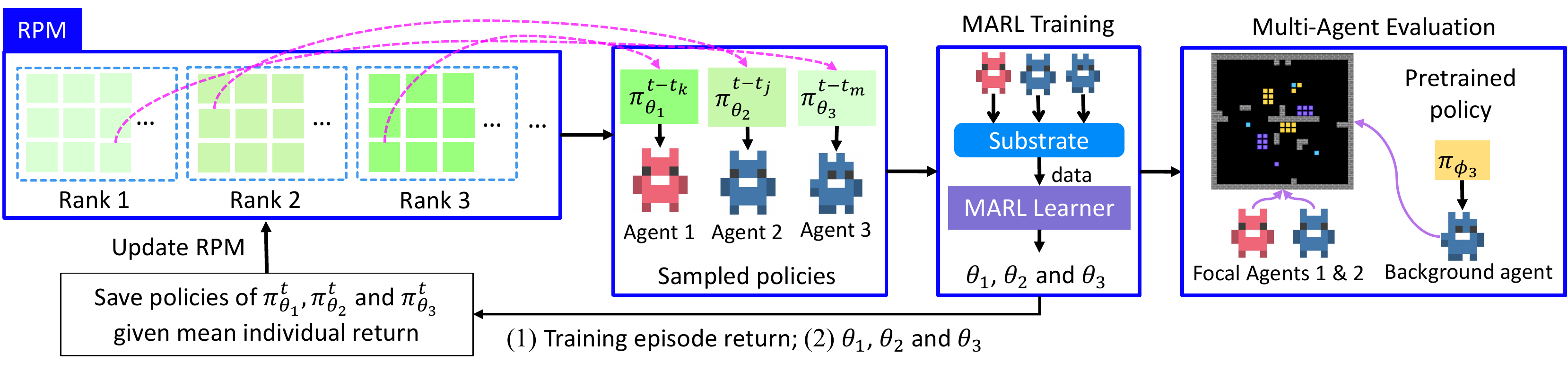}
        \caption{The workflow of RPM for  a three-agent substrate. In the workflow, there are 3 agents in the substrate. Agent 3 is the background agent. Agents 1 and 2 are focal agents.}\label{fig:rpm_arch}
    \end{center}
\end{figure}

\subsection{RPM: \underline{R}anked \underline{P}olicy \underline{M}emory}\label{sec:rpm}

In MARL, the focal agents need adaptively interact with background agents to complete given tasks. 
Formally, we define the objective for optimizing performance of the focal agents without exploiting their trajectories in the evaluation scenario for training the policies $\{\pi_{\theta_j}\}^{M}_{j=1}$:
\begin{equation}
    \max\mathcal{J}(\{\pi_{\theta_j}\}^{M}_{j=1}) \triangleq \max \E_{\rs_{0:\infty}\sim \rho^{0:\infty}_{\mathcal{G}^{\prime}}, a^{j}_{0:\infty}\sim \{\pi_{\theta_j}\}^{M}_{j=1} }\left[ \sum_{t=0}^{\infty}\gamma^{t} \frac{1}{M} \sum^{M}_{j=1} r^{j}_{t} \Bigg\lvert \mathcal{G}^{\prime} \right].
\end{equation}
To improve the generalization performance of MARL, it is crucial for agents in the substrate to cover as much as multi-agent interactions, \ie, data, that resemble the unseen multi-agent interactions in the evaluation scenario.
However, current training paradigms, like independent learning~\citep{tampuu2017multiagent} and centralized training and decentralized execution (CTDE)~\citep{oliehoek2008optimal}, cannot give diversified multi-agent interactions, as the agents' policies are trained at the same pace. 
To address this issue, we propose to gather massive diversified agent-agent interaction data for multi-agent learning. The diversified agent-agent interaction data are generated by agents that have different ranks of policies.
Concretely, we maintain a look-up memory during training, where each entry is a key and the corresponding value is a list. 
We take the training episode return of the agents' policies as the key, and use the list to store policies of the agents evaluated. 
When agents in the substrate start a new episode, there is a probability $p$ to replace all agents' behavior policies with the sampled policies from the memory.
This method is termed Ranked Policy Memory (RPM). 
Below we introduce how to build it and how to sample policies from it.

\textbf{RPM Building.}
We denote an RPM with $\rpmnotation$, which consists of $|R_{\text{max}}|$ entries, \ie, ranks, where $|R_{\text{max}}|$ is the maximum training episode return (the episode return in the substrate).
While agents are acting in the substrate, the training episode return $R$ of all agents (with policies $\{\pi^{i}_{\theta}\}^{N}_{i=1}$) is returned. Then $\{\pi^{i}_{\theta}\}^{N}_{i=1}$ are saved into $\rpmnotation$ by appending agents' policies into the corresponding memory slot, $\rpmnotation\texttt{[}r_e\texttt{]}.\texttt{add(}\{\pi^{i}_{e}\}^{N}_{i=1}\texttt{)}$.
To avoid there being too many entries in the policy memory caused by continuous episode return values, we discretize the training episode return.
The discretized entry $\kappa$ covers a range of $[\kappa, \kappa+\psi)$, where $\psi > 0$ and is an integer number. For the training episode return $R$, the corresponding entry $\kappa$ can be calculated by:
\begin{equation}
    \kappa = 
    \begin{cases}
          \floor{R / \psi} \times \bm{1}\{(R ~\texttt{mod}~\psi) \neq 0\} \times \psi, & \text{if}~R \geq 0,\\
          \floor{R / \psi} \times \psi, & \text{otherwise.}
    \end{cases}
\end{equation}
where $\bm{1}\{\cdot\}$ is the indicator function, and $\floor{\cdot}$ is the floor function. Intuitively, discretizing $R$ saves memory and memorize policies of similar performance in to the same rank. Therefore, diversified policies can be saved to be sampled for agents.

\textbf{RPM Sampling.} 
The memory $\rpmnotation$ stores diversified policies with different levels of performance.
We can sample various policies of different ranks and assign each policy to each agent in the substrate to collect multi-agent trajectories for training. 
These diversified multi-agent trajectories can resemble trajectories generated by the interaction with agents possessing unknown policies in the evaluation scenario.
At the beginning of an episode, we first randomly sample $N$ keys with replacement and then randomly sample one policy for each key from the corresponding list.
All agents' policies will be replaced with the newly sampled policies for multi-agent interactions in the substrate, thus generating diversified multi-agent trajectories.

\begin{wrapfigure}{R}{0.51\textwidth}
\begin{minipage}{0.51\textwidth}
\vspace{-4mm}
\IncMargin{1em}
\begin{algorithm2e}[H]
{
  \footnotesize
  \textbf{Input}: Initialize $\bm{\pi_\theta}$, $\rpmnotation$, $\mathcal{D}$, $\mathcal{G}$ and $\mathcal{G}^{\prime}$;\\
  \textbf{Input}: Initialize behavior policy $\bm{\pi}_{\bm{\theta}_{b}} \leftarrow \bm{\pi_\theta}$; \\
  \For{$t~\normalfont{\text{in}}~1, \dots, \normalfont{\texttt{MAX\_STEP}}$} {
    \uIf{\normalfont{\texttt{RPM sampling}}}{ \label{algline:check_sapmling}
        $\bm{\pi}_{\bm{\theta}_{b}} \leftarrow \texttt{SamplingRPM(}\rpmnotation\texttt{)}$; \label{algline:sapmling} \\
    }
    $\mathcal{D} \leftarrow \texttt{GatherTrajectories(}\bm{\pi}_{\bm{\theta}_{b}}, \mathcal{G}\texttt{)}$; \label{algline:gathering}\\
    $\bm{\pi_\theta} \leftarrow \texttt{MARLTrainig(}\bm{\pi_\theta}, \mathcal{D}\texttt{)}$;\label{algline:marltraining}\\
    $\rpmnotation \leftarrow \texttt{UpdateRPM(}\bm{\pi_\theta},\rpmnotation, \mathcal{G}\texttt{)}$; \label{algline:updaterpm}\\
    $\bar{R} \leftarrow \texttt{Evaluate(}\bm{\pi_\theta}, \mathcal{G}^{\prime}\texttt{)}$; \label{algline:eval_rpm}\\
    $\bm{\pi}_{\bm{\theta}_{b}} \leftarrow \bm{\pi_\theta}$; \label{algline:sync_behavior}\\
  }
  \textbf{Output}: $\bm{\pi_\theta}$.
  \caption{MARL with RPM} \label{algorithm:train_rpm}
}
\end{algorithm2e}
\vspace{-3mm}
\end{minipage}
\end{wrapfigure}
\textbf{MARL with RPM.} We showcase an example of the workflow of RPM in Figure~\ref{fig:rpm_arch}. There are three agents in training. Agents sample policies from RPM. Then all agents collect data in the substrate for training. The training episode return is then used to update RPM. During evaluation, agents 1 and 2 are selected as focal agents and agent 3 is selected as the background agent. 

We present the pseudo-code of MARL training with RPM in Algorithm~\ref{algorithm:train_rpm}. 
In Lines~\ref{algline:check_sapmling}-\ref{algline:sapmling}, the $\bm{\pi}_{\bm{\theta}_{b}}$ is updated by sampling policies from RPM. Then, new trajectories of $\mathcal{D}$ are collected in Line~\ref{algline:gathering}. $\bm{\pi_\theta}$ is trained in Line~\ref{algline:marltraining} with MARL method by using the newly collected trajecotries and $\bm{\pi}_{\bm{\theta}_{b}}$ is updated with the newly updated $\bm{\pi_\theta}$. RPM is updated in Line~\ref{algline:updaterpm}. After that, the performance of $\bm{\pi_\theta}$ is evaluated in the evaluation scenario $\mathcal{G}^{\prime}$ and the evaluation score $\bar{R}$ is returned in Line~\ref{algline:eval_rpm}.

\textbf{Discussion.} RPM leverages agents' previously trained models in substrates to cover as many patterns of multi-agent interactions as possible to achieve generalization of MARL agents when paired with agents with unseen policies in evaluation scenarios. It uses the self-play framework for data collection. Self-play~\citep{brown1951iterative,heinrich2015fictitious,silver2018general,baker2019emergent} maintains a memory of the opponent's previous policies for acquiring equilibria. RPM differs from other self-play methods in four aspects: (i) self-play utilizes agent's previous policies to create fictitious opponents when the real opponents are not available. By playing with the fictitious opponents, many fictitious data are generated for training the agents. In RPM, agents load their previous policies to diversify the multi-agent interactions, such as multi-agent coordination and social dilemmas, and all agents' policies are trained by utilizing the diversified multi-agent data.  (ii) Self-play does not maintain explicit ranks for policies while RPM maintains ranks of policies. (iii) Self-play was not introduced for generalization of MARL while RPM aims to improve the generalization of MARL. In Section~\ref{sec:experiments}, we also present the evaluation results of a self-play method.

\subsection{MARL Training}\label{sec:marl}

We incorporate RPM into the MARL training pipeline.
We take MAPPO~\citep{yu2021surprising} for instantiating our method, which is a multi-agent variant of PPO~\citep{schulman2017proximal} and outperforms many MARL methods~\citep{rashid2018qmix,rashid2020weighted,wang2021qplex} in various complex multi-agent domains.
In MAPPO, a central critic is maintained for utilizing the concealed information of agents to boost multi-agent learning due to non-stationarity. RPM introduces a novel method for agents to collect experiences/trajectories $\bm{\tau}=\{\tau_i\}^{N}_{i=1}$. Each agent optimizes the following objective:
\begin{equation}
    \mathcal{J}(\theta_i) = \mathbb{E} \left[ \texttt{min} \left(\eta^{t}_{i}\left(\theta^{t}_{i} \right) \cdot A^{t}_{i}, \texttt{clip}\left(\eta^{t}_{i}\left(\theta^{t}_{i}\right), 1-\epsilon, 1+\epsilon\right) \cdot A^{t}_{i} \right) \right],
\end{equation}
where $\eta^{t}_{i}(\theta^{t}_{i})=\frac{\pi_{\theta^{t}_{i}}(u^{t}_{i} | \tau^{t}_{i})}{\pi_{\theta^{\text{old}}_{i}}(u^{t}_{i}| \tau^{t}_{i})}$ denotes the important sampling weight. The $\texttt{clip}\left(\cdot\right)$ clips the values of $\theta^{i}$ that are outside the range $\left[ 1-\epsilon, 1+\epsilon \right]$ and $\epsilon$ is a hyperparameter. $A^{t}_{i}$ is a generalized advantage estimator (GAE)~\citep{schulman2015high}. To optimize the central critic $V_{\psi}(\{o^{t}_{i}, u^{t}_{i}\}^{N}_{i=1})$, we mix agents' observation-action pairs and output an $N$-head vector where each value corresponds to the agent's value:
\begin{equation}
    \mathcal{L} (\psi) :=\mathbb{E}_{\mathcal{D}^{\prime} \sim \mathcal{D}}\left[\left(y_{t}-V_{
    \bar{\psi}}(\{o^{t}_{i}, u^{t}_{i}\}^{N}_{i=1})\right)^{2}\right],
\end{equation}
where $y_{t}=\left[\sum_{l=0}^{k-1} \gamma^{l} r^{t+l}_{i}+\gamma^{k} V_{\bar{\psi}}(\{o^{t+k}_{i}, u^{t+k}_{i}\}^{N}_{i=1})\texttt{[}i\texttt{]}\right]^{N}_{i=1}$ is a vector of $k$-step returns, and $\mathcal{D}^{\prime}$ is a sample from the replay buffer $\mathcal{D}$. 
In complex scenarios, \textit{e.g.}, Melting Pot, 
with an agent's observation as input, its action would not impact other agents' return, since the global states contain redundant information that deteriorates multi-agent learning.
We present the whole training process, the network architectures of the agent and the central critic in Appendix~\ref{secapp:marl_training}.

\section{Experiments}\label{sec:experiments}

In this section, to verify the effectiveness of RPM in improving generalization of MARL, we conduct extensive experiments on Melting pot and present the empirical results. We first introduce Melting Pot, baselines and experiments setups. Then we present the main results to show the superiority of RPM. To demonstrate that $\psi$ is important for RPM, we conducted ablation studies. We finally showcase a case study to visualize RPM. To sum up, we answer the following questions:
\textbf{Q1}: Is RPM effective in boosting the generalization performance of MARL agents? \textbf{Q2}: Does the value of $\psi$ matter in RPM for training? \textbf{Q3}: Does RPM gather diversified policies and trajectories?

\begin{figure}
    \vspace{-0.6cm}
    \centering
    \includegraphics[scale=0.3855]{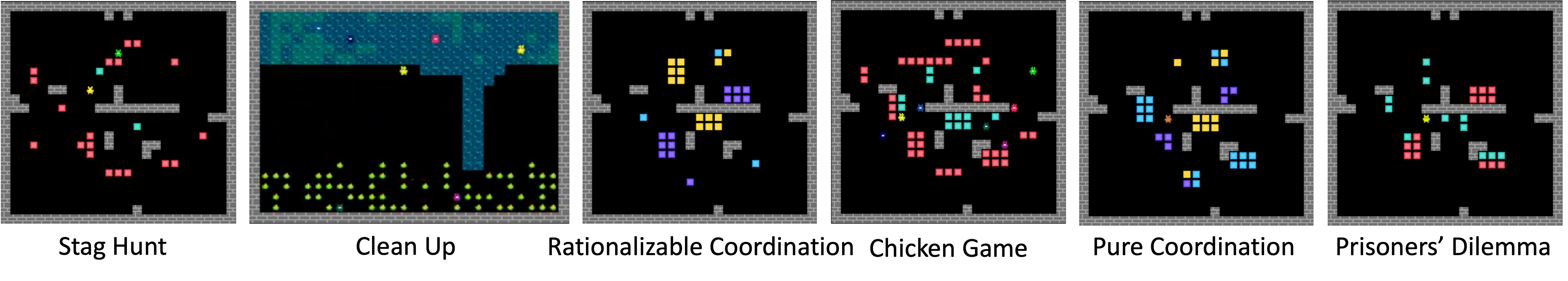}
    \vspace{-0.6cm}
    \caption{Melting Pot environments. More information can be found in Appendix~\ref{secapp:environments}.}
    \label{fig:mp_scenarios}
\end{figure}

\begin{table}
\centering
\caption{Properties of Melting Pot environments (substrates and evaluation scenarios). The first column shows the properties and the first row lists environments. \cmark~mark indicates the environment possessing the corresponding property while \xmark~mark stands for the environment that does not own the corresponding property. Refer to Appendix~\ref{secapp:environments} for more information about the environments.}
\vspace{-0.3cm}
\begin{tabular}{ccccccc}
\toprule
& \smaller{\textbf{Stag Hunt}} & \begin{tabular}{@{}c@{}}\smaller{\textbf{Pure}} \\ \smaller{\textbf{Coordination}}\end{tabular} & \smaller{\textbf{Clean Up}}  & \begin{tabular}{@{}c@{}}\smaller{\textbf{Prisoners'}} \\ \smaller{\textbf{Dilemma}}\end{tabular} &  \begin{tabular}{@{}c@{}}\smaller{\textbf{Rational}} \\ \smaller{\textbf{Coordination}}\end{tabular} & \smaller{\textbf{Chicken Game}} \\
\midrule
\smaller{\textbf{Temporal Coordination}} & \xmark & \xmark & \cmark & \xmark & \xmark & \xmark \\
\smaller{\textbf{Reciprocity}}           & \cmark & \cmark & \cmark & \cmark & \xmark & \cmark \\
\smaller{\textbf{Deception}}             & \cmark & \xmark & \cmark & \cmark & \xmark & \cmark \\
\smaller{\textbf{Fair Resource Sharing}} & \xmark & \xmark & \cmark & \xmark & \xmark & \xmark \\
\smaller{\textbf{Convention Following}}  & \cmark & \cmark & \cmark & \xmark & \cmark & \cmark \\
\smaller{\textbf{Task Partitioning}}     & \xmark & \xmark & \cmark & \cmark & \xmark & \xmark \\
\smaller{\textbf{Trust \& Partnership}}  & \cmark & \xmark & \xmark & \xmark & \xmark & \cmark \\
\smaller{\textbf{Free Riding}}           & \xmark & \xmark & \cmark & \xmark & \xmark & \xmark \\
\bottomrule
\end{tabular}
\label{tab:tb_mp_properties}
\end{table}

\subsection{Experimental Setup}

\begin{wrapfigure}{r}{0.42\textwidth}
\vspace{-0.7cm}
    \begin{center}
        \includegraphics[width=.42\textwidth]{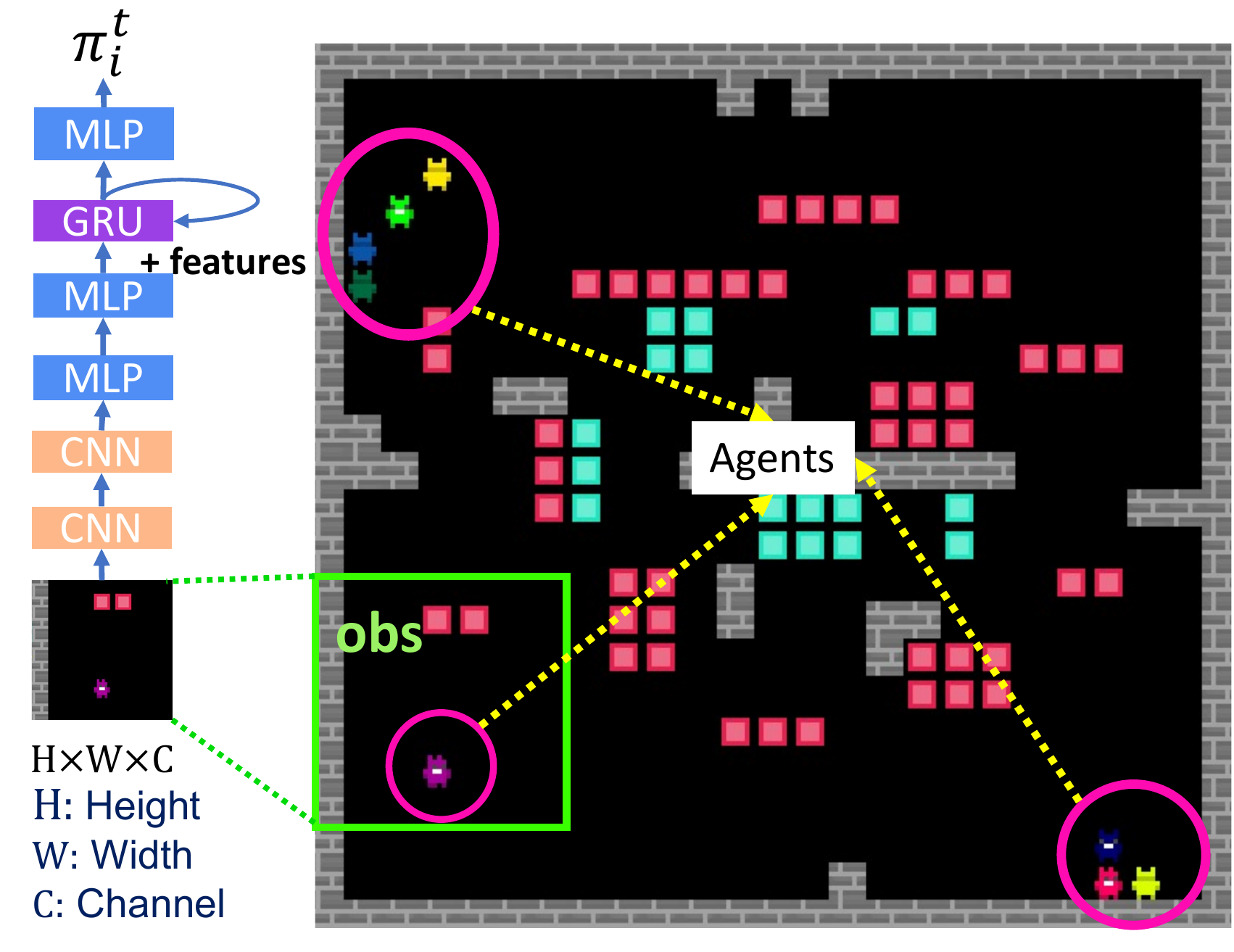}
        \vspace{-0.4cm}
        \caption{The green box to the lower left shows the agent's observation.}\label{fig:mp_example}
    \vspace{-0.3cm}
    \end{center}
\end{wrapfigure}
\textbf{Melting Pot.} To demonstrate that RPM enables MARL agents to learn generalizable behaviors, we carry out extensive experiments on DeepMind's Melting Pot~\citep{leibo2021scalable}. Melting Pot is a suite of testbeds for the generalization of MARL methods. It proposes a novel evaluation pipeline for the evaluation of the MARL method in various domains. That is, all MARL agents are trained in the substrate; during evaluation, some agents are selected as the focal agents (the agents to be evaluated) and the rest agents become the background agents (pretrained policies of MARL models will be plugged in); the evaluation scenarios share the same physical properties with the substrates. Melting Pot environments possess many properties, such as temporal coordination and free riding as depicted in Table~\ref{tab:tb_mp_properties}. MARL agent performing well in these environments means its behaviors demonstrate these properties.
In Figure~\ref{fig:mp_example}, the agent's observation is shown in the green box to the lower left of the state (\ie, the whole image). The agent is in the lower middle of the observation. The neural network architecture of the agent's policy is shown on the left. More information about the setting of substrates, neural network architectures, MARL training can be found in Appendix~\ref{secapp:marl_training}.

\textbf{Baselines.} Our baselines are MAPPO~\citep{yu2021surprising}, MAA2C~\citep{papoudakis2021benchmarking}, OPRE~\citep{vezhnevets2020options}, heuristic fictitious self-play (HFSP)~\citep{heinrich2017reinforcement,berner2019dota} and RandNet~\citep{lee2019network}. MAPPO and MAA2C are MARL methods that achieved outstanding performance in various multi-agent scenarios~\citep{papoudakis2021benchmarking}. OPRE was proposed for the generalization of MARL. RandNet is a general method for the generalization of RL by introducing a novel component in the convolutional neural network. HFSP is a general self-play method for obtaining equilibria in competitive games, we use it by using the policies saved by RPM.

\textbf{Training setup.} We use 6 representative substrates (Figure~\ref{fig:mp_scenarios}) to train MARL policies and choose one evaluation scenario from each substrate as our evaluation testbed. The properties of the environments are listed in~Table~\ref{tab:tb_mp_properties}. We train agents in Melting Pot substrates for 200 million frames with 3 random seeds for RPM and 4 seeds for baselines. Our training framework is a distributed framework where there are 30 CPU cores (actors) to collect experiences and 1 GPU for the learner to to learn policies. We implement our actors with Ray~\citep{moritz2018ray} and the learner with EPyMARL~\citep{papoudakis2021benchmarking}. We use mean-std to measure the performance of all methods. The bold lines in all figures are the mean and the shades stand for the standard deviation. Due to a limited computation budget, it is redundant in computation to compare our method with other methods such as QMIX~\citep{rashid2018qmix} and MADDPG~\citep{lowe2017multi} as MAPPO outperforms them. All experiments are conducted on NVIDIA A100 GPUs.

\begin{figure}
\vspace{-0.8cm}
\begin{center}
\includegraphics[scale=0.375]{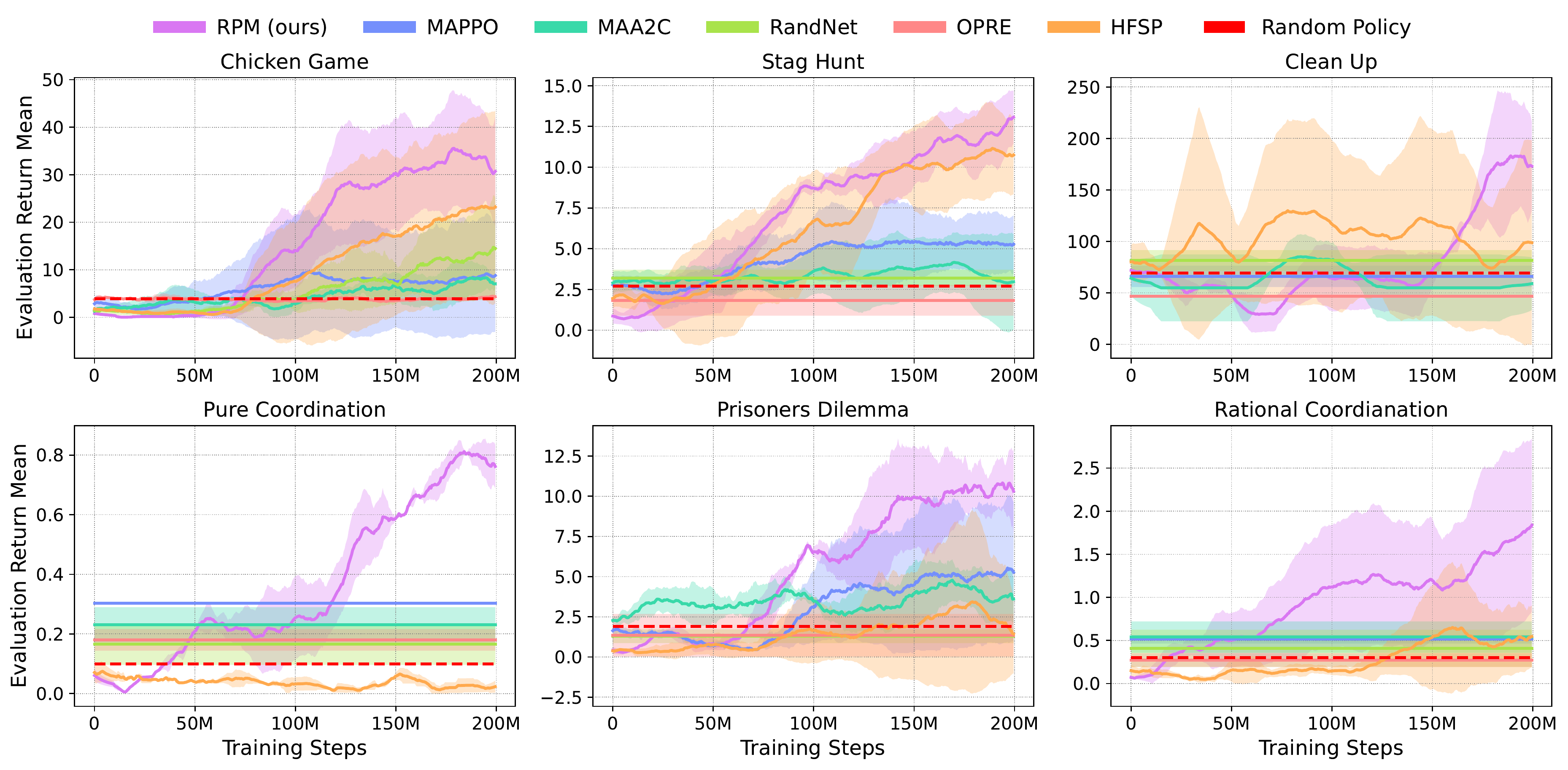}
\end{center}
\vspace{-.4cm}
\caption{Evaluation results of RPM and baselines on evaluation scenarios. The red dash horizontal lines indicate the results of random policy.}
\vspace{-0.4cm}
\label{fig:exp_mp_baselines}
\end{figure}

\subsection{Results}

To answer \textbf{Q1}, we present the evaluation results of 6 Melting Pot evaluation scenarios in ~Figure~\ref{fig:exp_mp_baselines}. We can find that our method can boost MARL in various evaluation scenarios which have different properties as shown in Table~\ref{tab:tb_mp_properties}. In Chicken Game (eval, `eval' stands for the evaluation scenario of the substrate Chicken Game), RPM outperforms its counterparts with a convincing margin. HFSP attains over 20 evaluation mean returns. RandNet gets around 15 evaluation mean returns. MAA2C and OPRE perform nearly random (the red dash lines indicate the random result). In Pure Coordination (eval), Rational Coordination (eval) and Prisoners' Dilemma (eval), most of the baselines perform poorly. In Stag Hunt (eval) and Clean Up (eval), MAPPO and MAA2C also perform unsatisfactorily. We can also find that HFSP even gets competitive performance in Stag Hunt (eval) and Clean Up (eval). However, HFSP performs poorly in Pure Coordination (eval), Rational Coordination (eval) and Prisoners' Dilemma (eval). Therefore, vanilla self-play method cannot directly be applied to improve the generalization of MARL methods. To sum up, RPM boosts the performance up to around \incres\% on average compared with MAPPO on 6 evaluation scenarios.

\subsection{Ablation Study}

\begin{wrapfigure}{r}{0.55\textwidth}
\vspace{-0.8cm}
\begin{center}
    \includegraphics[scale=0.45]{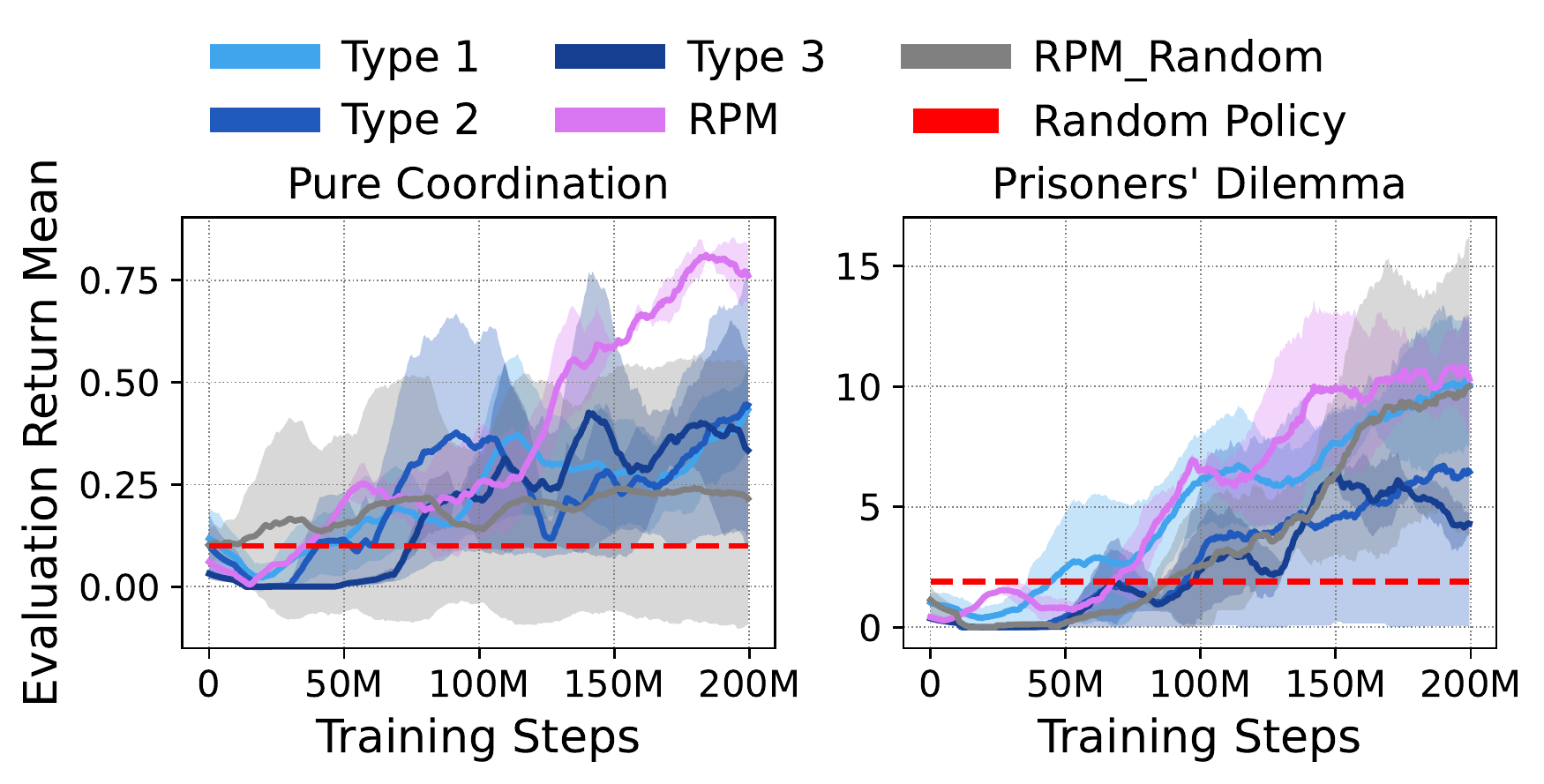}
\end{center}
\vspace{-0.5cm}
\caption{\footnotesize{Ablation Studies: the performance of RPM with 3 types of $\psi$ and Random sampling (without ranks).}}
\vspace{-0.3cm}
\label{fig:exp_mp_ablation_3_app}
\end{wrapfigure}
To investigate which value of $\psi$ has the greatest impact on RPM performance, we conduct ablation studies by (i) removing ranks and sampling from the checkpoint directly; (ii) reducing the number of ranks by changing the value of $\psi$. As shown in Figure~\ref{fig:exp_mp_ablation_3_app}, without ranks (sampling policies without ranks randomly), RPM cannot perform well in all evaluation scenarios. Especially in Pure Coordination (eval) where the result is low and has large variance. In RPM, choosing the right interval $\psi$ can improve the performance as shown in the results of Pure Coordination (eval) and Prisoners' Dilemma (eval), showing that the value of $\psi$ is important for RPM. We summarize the results and values of $\psi$ in Table~\ref{table:ablation_studies_1} and Table~\ref{table:ablation_studies_2}.

\begin{minipage}{\textwidth}
    \begin{minipage}[b]{0.55\textwidth}
    \setlength{\tabcolsep}{1.8mm}
    \centering
    \smaller
    \captionof{table}{\footnotesize{Ablation study: the final mean evaluation episode return. Curves are in Figure~\ref{fig:exp_mp_ablation_3_app}.}}
    \vspace{-2mm}
    \begin{tabular}{@{}rrrrrr@{}}
        \toprule
        \multirow{2}{*}{\textbf{Eval Scenarios}} &
        \multirow{2}{*}{\textbf{RPM}} &
        \multirow{2}{*}{\textbf{Random}} &
        \multicolumn{3}{c}{\textbf{Types of} $\psi$}
        \\
        \cmidrule(l){4-6}
        & & & $1$ & ${2}$ & ${3}$ \\
        \midrule
        \setcounter{MinX}{20}%
        \setcounter{MidX}{35}%
        \setcounter{MaxX}{54}%
        Pure Coordination & \ApplyPositiveGradient{0.76} & \ApplyPositiveGradient{0.22} & \ApplyPositiveGradient{0.40} & \ApplyPositiveGradient{0.42} & \ApplyPositiveGradient{0.36} \\
        \setcounter{MinX}{420}%
        \setcounter{MidX}{700}%
        \setcounter{MaxX}{1060}%
        Prisoners' Dilemma & \ApplyPositiveGradient{10.56} & \ApplyPositiveGradient{9.78} & \ApplyPositiveGradient{10.0} & \ApplyPositiveGradient{6.35} & \ApplyPositiveGradient{4.24} \\
        \bottomrule
    \end{tabular}
    \label{table:ablation_studies_1}
  \end{minipage}
  \hfill
  \begin{minipage}[b]{0.44\textwidth}
    \smaller
    \centering
    \captionof{table}{\footnotesize{$\psi$ values. $\psi^{*}$ indicates the values of $\psi$ used to get results in Figure~\ref{fig:exp_mp_baselines}.}}
    \vspace{-2mm}
    \begin{tabular}{@{}rrrrr@{}}
        \toprule
        \multirow{2}{*}{\textbf{Eval Scenarios}} &
        \multirow{2}{*}{\textbf{$\psi^{*}$}} &
        \multicolumn{3}{c}{\textbf{Types of} $\psi$}
        \\
        \cmidrule(l){3-5}
        & & $1$ & ${2}$ & ${3}$ \\
        \midrule
        \rule{0pt}{8.7pt}
        Pure Coordination & 0.01 & 0.1 & 0.5 & 1 \\
        \rule{0pt}{8.7pt}
        Prisoners' Dilemma & 0.02 & 0.2 & 1 & 5 \\
        \bottomrule
    \end{tabular}
    \label{table:ablation_studies_2}
  \end{minipage}
\end{minipage}

\subsection{Case Study}

\begin{figure}
\vspace{-0.6cm}
\begin{center}
    \includegraphics[scale=0.50]{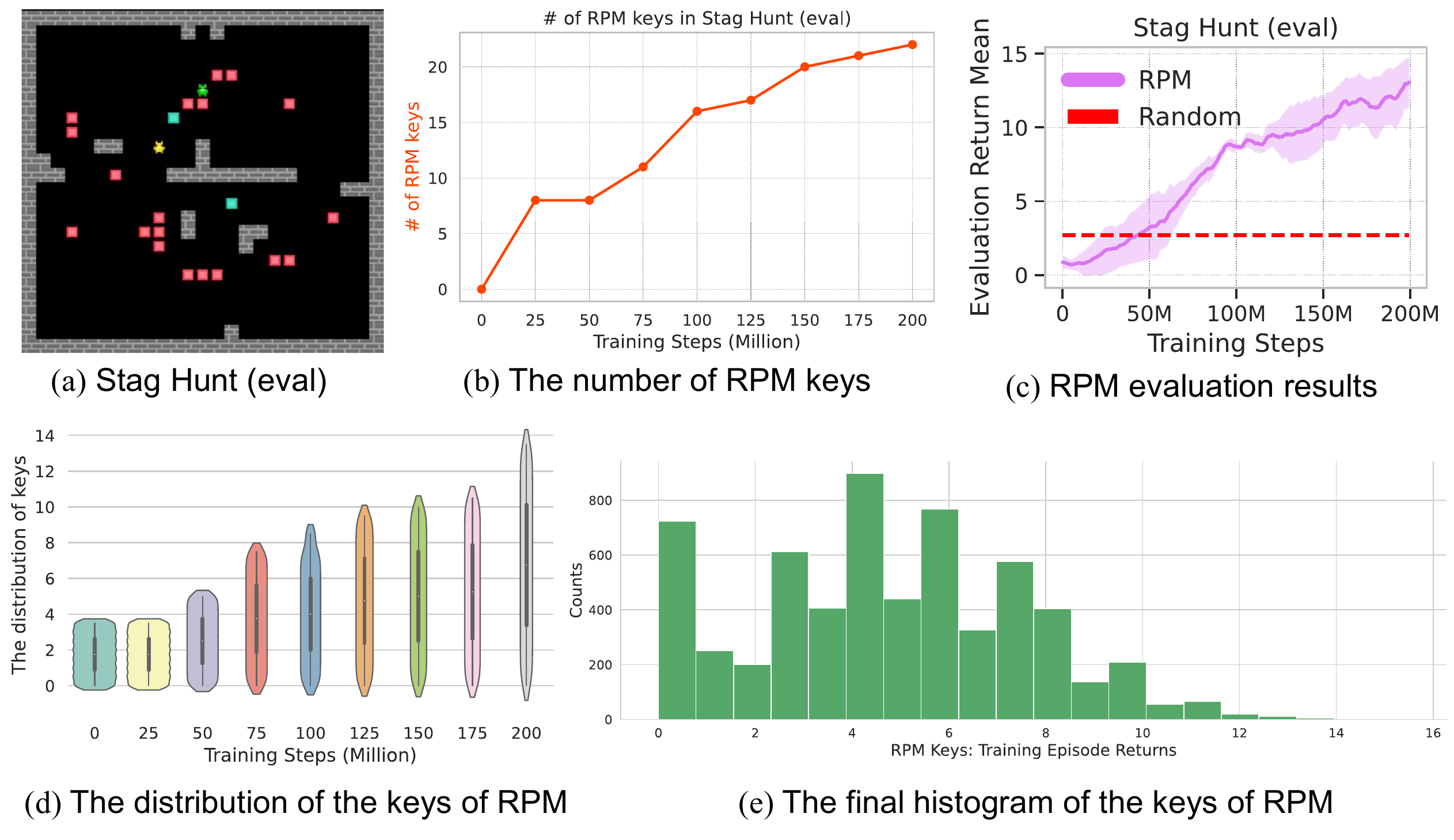}
\end{center}
\vspace{-0.4cm}
\caption{Result analysis. \textbf{(a)} The pictorial view of Stag Hunt (eval); \textbf{(b)} The number of RPM keys during training; \textbf{(c)} The evaluation results of RPM on Stag Hunt (eval); \textbf{(d)} The distribution of the keys of RPM during training; \textbf{(e)} The histogram of the keys of RPM at timestep 200M during training.}
\vspace{-0.4cm}
\label{fig:exp_mp_stag_hunt_analysis}
\end{figure}

We showcase how RPM helps to train the focal agents to choose the right behaviors in the evaluation scenario after training in the substrate. To illustrate the trained performance of RPM agents, we use the RPM agent trained on Stag Hunt and run the evaluation on Stag Hunt (eval). In Stag Hunt, there are 8 agents in this environment. Each agent collects resources that represent `hare' (red) or `stag' (green) and compares inventories in an interaction, \ie, encounter. The results of solving the encounter are the same as the classic Stag Hunt matrix game. In this environment, agents are facing tension between the reward for the team and the risk for the individual. In Stag Hunt (eval) (Figure~\ref{fig:exp_mp_stag_hunt_analysis} (a)). One focal agent interacts with seven pretrained agents. All background agents were trained to play the `stag' strategy during the interaction\footnote{This preference was trained with pseudo rewards by~\cite{leibo2021scalable} and the trained models are available at this link: \url{https://github.com/deepmind/meltingpot}}. The optimal policy for the focal agent is also to play `stag'. However, it is challenging for agents to detect other agents' strategy since such a behavior may not persist in the substrate. Luckily, RPM enables focal agents to behave correctly in this scenario.

To answer \textbf{Q3}, we present the analysis of RPM on the substrate Stag Hunt and its evaluation scenario Stag Hunt (eval) in Figure~\ref{fig:exp_mp_stag_hunt_analysis}. We can find that in Figure~\ref{fig:exp_mp_stag_hunt_analysis} (b), the number of the keys in RPM is growing monotonically during training and the maximum number of the keys in RPM is over 20, showing that agents trained with RPM discover many novel patterns of multi-agent interaction and new keys are created and the trained models are saved in RPM. Meanwhile, the evaluation performance is also increasing in Stag Hunt (eval) as depicted in ~Figure~\ref{fig:exp_mp_stag_hunt_analysis} (c). In ~Figure~\ref{fig:exp_mp_stag_hunt_analysis} (d), it is interesting to see that the distribution of the keys of RPM is expanding during training. In the last 25 million training steps, the last distribution of RPM keys covers all policies of different levels of performance, ranging from 0 to 14. By utilizing RPM, agents can collect diversified multi-agent trajectories for multi-agent training. ~Figure~\ref{fig:exp_mp_stag_hunt_analysis} (e) demonstrates the final histogram of RPM keys after training. There are over 600 trained policies that have small value of keys. Since at the early stage of training, agents should explore the environment, it is reasonable to find that a large number of trained policies of RPM keys have low training episode returns. After 50 million training steps, RPM has more policies that have higher training episode returns. Note that the maximum training episode return of RPM keys is over 14 while the maximum mean evaluation return of RPM shown in ~Figure~\ref{fig:exp_mp_stag_hunt_analysis} (c) is around 14.

In our experiments, we find that training policies with good performance in the substrate is crucial for improving generalization performance in the evaluation scenarios. When MARL agents perform poorly, \ie, execute sub-optimal actions, the evaluation performance will be also inferior or even random, making it is hard to have diversified policies. We show the results in Appendix~\ref{secapp:results}.

\section{Related Works}\label{sec:relatedworks}
Recent advances in MARL~\citep{yang2020overview,zhang2021multi} have demonstrated its success in various complex multi-agent domains, including multi-agent coordination~\citep{lowe2017multi,rashid2018qmix,wang2021dop}, real-time strategy~(RTS) games~\citep{jaderberg2019human,berner2019dota,vinyals2019grandmaster}, social dilemma~\citep{leibo2017multi,wang2018towards,jaques2019social,vezhnevets2020options}, multi-agent communication~\citep{foerster2016learning,yuan2022multi}, asynchronous multi-agent learning~\citep{amato2019modeling,qiu2022off}, open-ended environment~\citep{team2021open}, autonomous systems~\citep{huttenrauch2017guided,peng2021learning} and game theory equilibrium solving~\citep{lanctot2017unified,perolat2022mastering}. Despite strides made in MARL, training generalizable behaviors in MARL is yet to be investigated. 
Generalization in RL~\citep{packer2018assessing,song2019observational,ghosh2021generalization,lyle2022learning} has achieved much progress in domain adaptation~\citep{higgins2017darla} and procedurally generated environments~\citep{lee2019network,igl2020transient,zha2020rank} in recent years. However, there are few works of generalization in MARL domains~\citep{,carion2019structured,vezhnevets2020options,mahajan2022generalization,mckee2022quantifying}.
Recently, \cite{vezhnevets2020options} propose a hierarchical MARL method for agents to play against opponents
it hasn’t seen during training. However, the evaluation scenarios are only limited to  simple competitive scenarios. \cite{mahajan2022generalization} investigated the generalization in MARL empirically and proposed theoretical findings based on successor features~\citep{dayan1993improving,barreto2018transfer}. However, there is no method proposed to achieve generalization in MARL.

Ad-hoc team building~\citep{stone2010teach,gu2021online} models the multi-agent problem as a single-agent learning task. In ad-hoc team building, one ad-hoc agent is trained by interacting with agents that have fixed pretrained policies and the  non-stationarity issue is not severe. However, in our formulation, non-stationarity is the main obstacle to MARL training. In addition, there is only one ad-hoc agent evaluated by interacting agents that are unseen during training while there can be more than one focal agent in our formulation as defined in  Definition~\ref{defintion:multi_agent_evaluation}, thus making our formulation general and challenging. There has been a growing interest in applying self-play to solve complex games~\citep{heinrich2015fictitious,silver2018general,hernandez2019generalized,baker2019emergent}; however, its value in enhancing the generalization of MARL agents has yet to be examined.

\section{Conclusion, Limitations and Future Work}\label{sec:conclusion}

In this paper, we consider the problem of achieving generalizable behaviors in MARL. We first model social learning with Markov Game.
In order to train agents that are able to interact with agents that possess unseen policies.
We propose a simple yet effective method, RPM, to save policies of different levels. We save policies by ranking the training episode return. Empirically, RPM significantly boosts the performance of MARL agents in a variety of Melting Pot evaluation scenarios.

RPM's performance is highly dependent on the appropriate value of $\psi$. Several attempts are required to determine the correct value of $\psi$ for RPM. We are interested in discovering more broad measures for ranking policies that do not explicitly consider the training episode return. Recently, there is a growing interest in planning in RL, especially with model-based RL. We are interested in exploring the direction of applying planning and opponent/teammate modelling for attaining generalized behaviors with MARL for future work. In multi-agent scenarios, agents are engaged in complex interactions. Devising novel self-play method is our future direction for improving generalization of MARL methods.

\newpage
\section{Ethics Statement}

We addressed the relevant aspects in our conclusion and have no conflicts of interest to declare.

\section{Reproducibility Statement}

We provide detailed descriptions of our experiments in the appendix and list all relevant parameters in Table~\ref{tab:hyparam} and Table~\ref{tab:psi_table} in Appendix~\ref{secapp:marl_training}. The code can be found at this anonymous link: \url{https://sites.google.com/view/rpm-2022/}.

\clearpage
\bibliography{main}

\begin{thebibliography}{69}
\providecommand{\natexlab}[1]{#1}
\providecommand{\url}[1]{\texttt{#1}}
\expandafter\ifx\csname urlstyle\endcsname\relax
  \providecommand{\doi}[1]{doi: #1}\else
  \providecommand{\doi}{doi: \begingroup \urlstyle{rm}\Url}\fi

\bibitem[Amato et~al.(2019)Amato, Konidaris, Kaelbling, and
  How]{amato2019modeling}
Christopher Amato, George Konidaris, Leslie~P Kaelbling, and Jonathan~P How.
\newblock Modeling and planning with macro-actions in decentralized pomdps.
\newblock \emph{Journal of Artificial Intelligence Research}, 64:\penalty0
  817--859, 2019.

\bibitem[Bacon et~al.(2017)Bacon, Harb, and Precup]{bacon2017option}
Pierre-Luc Bacon, Jean Harb, and Doina Precup.
\newblock The option-critic architecture.
\newblock In \emph{Proceedings of the AAAI Conference on Artificial
  Intelligence}, volume~31, 2017.

\bibitem[Baker et~al.(2019)Baker, Kanitscheider, Markov, Wu, Powell, McGrew,
  and Mordatch]{baker2019emergent}
Bowen Baker, Ingmar Kanitscheider, Todor Markov, Yi~Wu, Glenn Powell, Bob
  McGrew, and Igor Mordatch.
\newblock Emergent tool use from multi-agent autocurricula.
\newblock In \emph{International Conference on Learning Representations}, 2019.

\bibitem[Barreto et~al.(2018)Barreto, Borsa, Quan, Schaul, Silver, Hessel,
  Mankowitz, Zidek, and Munos]{barreto2018transfer}
Andre Barreto, Diana Borsa, John Quan, Tom Schaul, David Silver, Matteo Hessel,
  Daniel Mankowitz, Augustin Zidek, and Remi Munos.
\newblock Transfer in deep reinforcement learning using successor features and
  generalised policy improvement.
\newblock In \emph{International Conference on Machine Learning}, pp.\
  501--510. PMLR, 2018.

\bibitem[Berner et~al.(2019)Berner, Brockman, Chan, Cheung, D{\k{e}}biak,
  Dennison, Farhi, Fischer, Hashme, Hesse, et~al.]{berner2019dota}
Christopher Berner, Greg Brockman, Brooke Chan, Vicki Cheung, Przemys{\l}aw
  D{\k{e}}biak, Christy Dennison, David Farhi, Quirin Fischer, Shariq Hashme,
  Chris Hesse, et~al.
\newblock Dota 2 with large scale deep reinforcement learning.
\newblock \emph{arXiv preprint arXiv:1912.06680}, 2019.

\bibitem[Brown(1951)]{brown1951iterative}
George~W Brown.
\newblock Iterative solution of games by fictitious play.
\newblock \emph{Act. Anal. Prod Allocation}, 13\penalty0 (1):\penalty0 374,
  1951.

\bibitem[Carion et~al.(2019)Carion, Usunier, Synnaeve, and
  Lazaric]{carion2019structured}
Nicolas Carion, Nicolas Usunier, Gabriel Synnaeve, and Alessandro Lazaric.
\newblock A structured prediction approach for generalization in cooperative
  multi-agent reinforcement learning.
\newblock \emph{Advances in Neural Information Processing Systems}, 32, 2019.

\bibitem[Cho et~al.(2014)Cho, Van~Merri{\"e}nboer, Bahdanau, and
  Bengio]{cho2014properties}
Kyunghyun Cho, Bart Van~Merri{\"e}nboer, Dzmitry Bahdanau, and Yoshua Bengio.
\newblock On the properties of neural machine translation: Encoder-decoder
  approaches.
\newblock \emph{arXiv preprint arXiv:1409.1259}, 2014.

\bibitem[Dayan(1993)]{dayan1993improving}
Peter Dayan.
\newblock Improving generalization for temporal difference learning: The
  successor representation.
\newblock \emph{Neural Computation}, 5\penalty0 (4):\penalty0 613--624, 1993.

\bibitem[Espeholt et~al.(2018)Espeholt, Soyer, Munos, Simonyan, Mnih, Ward,
  Doron, Firoiu, Harley, Dunning, et~al.]{espeholt2018impala}
Lasse Espeholt, Hubert Soyer, Remi Munos, Karen Simonyan, Vlad Mnih, Tom Ward,
  Yotam Doron, Vlad Firoiu, Tim Harley, Iain Dunning, et~al.
\newblock Impala: Scalable distributed deep-rl with importance weighted
  actor-learner architectures.
\newblock In \emph{International Conference on Machine Learning}, pp.\
  1407--1416. PMLR, 2018.

\bibitem[Foerster et~al.(2016)Foerster, Assael, de~Freitas, and
  Whiteson]{foerster2016learning}
Jakob Foerster, Ioannis~Alexandros Assael, Nando de~Freitas, and Shimon
  Whiteson.
\newblock Learning to communicate with deep multi-agent reinforcement learning.
\newblock In \emph{Advances in Neural Information Processing Systems}, pp.\
  2137--2145, 2016.

\bibitem[Fujimoto et~al.(2018)Fujimoto, Hoof, and
  Meger]{fujimoto2018addressing}
Scott Fujimoto, Herke Hoof, and David Meger.
\newblock Addressing function approximation error in actor-critic methods.
\newblock In \emph{International conference on machine learning}, pp.\
  1587--1596. PMLR, 2018.

\bibitem[Ghosh et~al.(2021)Ghosh, Rahme, Kumar, Zhang, Adams, and
  Levine]{ghosh2021generalization}
Dibya Ghosh, Jad Rahme, Aviral Kumar, Amy Zhang, Ryan~P Adams, and Sergey
  Levine.
\newblock Why generalization in rl is difficult: Epistemic pomdps and implicit
  partial observability.
\newblock \emph{Advances in Neural Information Processing Systems}, 34, 2021.

\bibitem[Gu et~al.(2021)Gu, Zhao, Hao, and An]{gu2021online}
Pengjie Gu, Mengchen Zhao, Jianye Hao, and Bo~An.
\newblock Online ad hoc teamwork under partial observability.
\newblock In \emph{International Conference on Learning Representations}, 2021.

\bibitem[Heinrich(2017)]{heinrich2017reinforcement}
Johannes Heinrich.
\newblock \emph{Reinforcement learning from self-play in imperfect-information
  games}.
\newblock PhD thesis, UCL (University College London), 2017.

\bibitem[Heinrich et~al.(2015)Heinrich, Lanctot, and
  Silver]{heinrich2015fictitious}
Johannes Heinrich, Marc Lanctot, and David Silver.
\newblock Fictitious self-play in extensive-form games.
\newblock In \emph{International conference on machine learning}, pp.\
  805--813. PMLR, 2015.

\bibitem[Hernandez et~al.(2019)Hernandez, Denamgana{\"\i}, Gao, York, Devlin,
  Samothrakis, and Walker]{hernandez2019generalized}
Daniel Hernandez, Kevin Denamgana{\"\i}, Yuan Gao, Peter York, Sam Devlin,
  Spyridon Samothrakis, and James~Alfred Walker.
\newblock A generalized framework for self-play training.
\newblock In \emph{2019 IEEE Conference on Games (CoG)}, pp.\  1--8. IEEE,
  2019.

\bibitem[Higgins et~al.(2017)Higgins, Pal, Rusu, Matthey, Burgess, Pritzel,
  Botvinick, Blundell, and Lerchner]{higgins2017darla}
Irina Higgins, Arka Pal, Andrei Rusu, Loic Matthey, Christopher Burgess,
  Alexander Pritzel, Matthew Botvinick, Charles Blundell, and Alexander
  Lerchner.
\newblock Darla: Improving zero-shot transfer in reinforcement learning.
\newblock In \emph{International Conference on Machine Learning}, pp.\
  1480--1490. PMLR, 2017.

\bibitem[Hochreiter \& Schmidhuber(1997)Hochreiter and
  Schmidhuber]{hochreiter1997long}
Sepp Hochreiter and J{\"u}rgen Schmidhuber.
\newblock Long short-term memory.
\newblock \emph{Neural computation}, 9\penalty0 (8):\penalty0 1735--1780, 1997.

\bibitem[Hupkes et~al.(2020)Hupkes, Dankers, Mul, and
  Bruni]{hupkes2020compositionality}
Dieuwke Hupkes, Verna Dankers, Mathijs Mul, and Elia Bruni.
\newblock Compositionality decomposed: how do neural networks generalise?
\newblock \emph{Journal of Artificial Intelligence Research}, 67:\penalty0
  757--795, 2020.

\bibitem[H{\"u}ttenrauch et~al.(2017)H{\"u}ttenrauch, {\v{S}}o{\v{s}}i{\'c},
  and Neumann]{huttenrauch2017guided}
Maximilian H{\"u}ttenrauch, Adrian {\v{S}}o{\v{s}}i{\'c}, and Gerhard Neumann.
\newblock Guided deep reinforcement learning for swarm systems.
\newblock \emph{In AAMAS 2017 Autonomous Robots and Multirobot Systems (ARMS)
  Workshop}, 2017.

\bibitem[Igl et~al.(2020)Igl, Farquhar, Luketina, Boehmer, and
  Whiteson]{igl2020transient}
Maximilian Igl, Gregory Farquhar, Jelena Luketina, Wendelin Boehmer, and Shimon
  Whiteson.
\newblock Transient non-stationarity and generalisation in deep reinforcement
  learning.
\newblock In \emph{International Conference on Learning Representations}, 2020.

\bibitem[Jaderberg et~al.(2019)Jaderberg, Czarnecki, Dunning, Marris, Lever,
  Castaneda, Beattie, Rabinowitz, Morcos, Ruderman, et~al.]{jaderberg2019human}
Max Jaderberg, Wojciech~M Czarnecki, Iain Dunning, Luke Marris, Guy Lever,
  Antonio~Garcia Castaneda, Charles Beattie, Neil~C Rabinowitz, Ari~S Morcos,
  Avraham Ruderman, et~al.
\newblock Human-level performance in 3{D} multiplayer games with
  population-based reinforcement learning.
\newblock \emph{Science}, 364\penalty0 (6443):\penalty0 859--865, 2019.

\bibitem[Jaques et~al.(2019)Jaques, Lazaridou, Hughes, Gulcehre, Ortega,
  Strouse, Leibo, and De~Freitas]{jaques2019social}
Natasha Jaques, Angeliki Lazaridou, Edward Hughes, Caglar Gulcehre, Pedro
  Ortega, DJ~Strouse, Joel~Z Leibo, and Nando De~Freitas.
\newblock Social influence as intrinsic motivation for multi-agent deep
  reinforcement learning.
\newblock In \emph{International Conference on Machine Learning}, pp.\
  3040--3049. PMLR, 2019.

\bibitem[Kirk et~al.(2021)Kirk, Zhang, Grefenstette, and
  Rockt{\"a}schel]{kirk2021survey}
Robert Kirk, Amy Zhang, Edward Grefenstette, and Tim Rockt{\"a}schel.
\newblock A survey of generalisation in deep reinforcement learning.
\newblock \emph{arXiv preprint arXiv:2111.09794}, 2021.

\bibitem[Lanctot et~al.(2017)Lanctot, Zambaldi, Gruslys, Lazaridou, Tuyls,
  P{\'e}rolat, Silver, and Graepel]{lanctot2017unified}
Marc Lanctot, Vinicius Zambaldi, Audrunas Gruslys, Angeliki Lazaridou, Karl
  Tuyls, Julien P{\'e}rolat, David Silver, and Thore Graepel.
\newblock A unified game-theoretic approach to multiagent reinforcement
  learning.
\newblock \emph{Advances in Neural Information Processing Systems}, 30, 2017.

\bibitem[Lee et~al.(2019)Lee, Lee, Shin, and Lee]{lee2019network}
Kimin Lee, Kibok Lee, Jinwoo Shin, and Honglak Lee.
\newblock Network randomization: A simple technique for generalization in deep
  reinforcement learning.
\newblock In \emph{International Conference on Learning Representations}, 2019.

\bibitem[Leibo et~al.(2017)Leibo, Zambaldi, Lanctot, Marecki, and
  Graepel]{leibo2017multi}
Joel~Z Leibo, Vinicius Zambaldi, Marc Lanctot, Janusz Marecki, and Thore
  Graepel.
\newblock Multi-agent reinforcement learning in sequential social dilemmas.
\newblock \emph{arXiv preprint arXiv:1702.03037}, 2017.

\bibitem[Leibo et~al.(2021)Leibo, Due{\~n}ez-Guzman, Vezhnevets, Agapiou,
  Sunehag, Koster, Matyas, Beattie, Mordatch, and Graepel]{leibo2021scalable}
Joel~Z Leibo, Edgar~A Due{\~n}ez-Guzman, Alexander Vezhnevets, John~P Agapiou,
  Peter Sunehag, Raphael Koster, Jayd Matyas, Charlie Beattie, Igor Mordatch,
  and Thore Graepel.
\newblock Scalable evaluation of multi-agent reinforcement learning with
  melting pot.
\newblock In \emph{International Conference on Machine Learning}, pp.\
  6187--6199. PMLR, 2021.

\bibitem[Lillicrap et~al.(2016)Lillicrap, Hunt, Pritzel, Heess, Erez, Tassa,
  Silver, and Wierstra]{lillicrap2015continuous}
Timothy~P Lillicrap, Jonathan~J Hunt, Alexander Pritzel, Nicolas Heess, Tom
  Erez, Yuval Tassa, David Silver, and Daan Wierstra.
\newblock Continuous control with deep reinforcement learning.
\newblock In \emph{International Conference on Learning Representations}, 2016.

\bibitem[Littman(1994)]{littman1994markov}
Michael~L Littman.
\newblock Markov games as a framework for multi-agent reinforcement learning.
\newblock In \emph{Machine learning proceedings 1994}, pp.\  157--163.
  Elsevier, 1994.

\bibitem[Lowe et~al.(2017)Lowe, Wu, Tamar, Harb, Abbeel, and
  Mordatch]{lowe2017multi}
Ryan Lowe, Yi~Wu, Aviv Tamar, Jean Harb, OpenAI~Pieter Abbeel, and Igor
  Mordatch.
\newblock Multi-agent actor-critic for mixed cooperative-competitive
  environments.
\newblock In \emph{Advances in Neural Information Processing Systems}, pp.\
  6379--6390, 2017.

\bibitem[Lyle et~al.(2022)Lyle, Rowland, Dabney, Kwiatkowska, and
  Gal]{lyle2022learning}
Clare Lyle, Mark Rowland, Will Dabney, Marta Kwiatkowska, and Yarin Gal.
\newblock Learning dynamics and generalization in reinforcement learning.
\newblock \emph{arXiv preprint arXiv:2206.02126}, 2022.

\bibitem[Mahajan et~al.(2022)Mahajan, Samvelyan, Gupta, Ellis, Sun,
  Rockt{\"a}schel, and Whiteson]{mahajan2022generalization}
Anuj Mahajan, Mikayel Samvelyan, Tarun Gupta, Benjamin Ellis, Mingfei Sun, Tim
  Rockt{\"a}schel, and Shimon Whiteson.
\newblock Generalization in cooperative multi-agent systems.
\newblock \emph{arXiv preprint arXiv:2202.00104}, 2022.

\bibitem[McKee et~al.(2022)McKee, Leibo, Beattie, and
  Everett]{mckee2022quantifying}
Kevin~R McKee, Joel~Z Leibo, Charlie Beattie, and Richard Everett.
\newblock Quantifying the effects of environment and population diversity in
  multi-agent reinforcement learning.
\newblock \emph{Autonomous Agents and Multi-Agent Systems}, 36\penalty0
  (1):\penalty0 1--16, 2022.

\bibitem[Mnih et~al.(2015)Mnih, Kavukcuoglu, Silver, Rusu, Veness, Bellemare,
  Graves, Riedmiller, Fidjeland, Ostrovski, et~al.]{mnih2015human}
Volodymyr Mnih, Koray Kavukcuoglu, David Silver, Andrei~A Rusu, Joel Veness,
  Marc~G Bellemare, Alex Graves, Martin Riedmiller, Andreas~K Fidjeland, Georg
  Ostrovski, et~al.
\newblock Human-level control through deep reinforcement learning.
\newblock \emph{Nature}, 518\penalty0 (7540):\penalty0 529--533, 2015.

\bibitem[Mnih et~al.(2016)Mnih, Badia, Mirza, Graves, Lillicrap, Harley,
  Silver, and Kavukcuoglu]{mnih2016asynchronous}
Volodymyr Mnih, Adria~Puigdomenech Badia, Mehdi Mirza, Alex Graves, Timothy
  Lillicrap, Tim Harley, David Silver, and Koray Kavukcuoglu.
\newblock Asynchronous methods for deep reinforcement learning.
\newblock In \emph{International conference on machine learning}, pp.\
  1928--1937. PMLR, 2016.

\bibitem[Moritz et~al.(2018)Moritz, Nishihara, Wang, Tumanov, Liaw, Liang,
  Elibol, Yang, Paul, Jordan, et~al.]{moritz2018ray}
Philipp Moritz, Robert Nishihara, Stephanie Wang, Alexey Tumanov, Richard Liaw,
  Eric Liang, Melih Elibol, Zongheng Yang, William Paul, Michael~I Jordan,
  et~al.
\newblock Ray: A distributed framework for emerging $\{$AI$\}$ applications.
\newblock In \emph{13th USENIX Symposium on Operating Systems Design and
  Implementation (OSDI 18)}, pp.\  561--577, 2018.

\bibitem[Oliehoek et~al.(2008)Oliehoek, Spaan, and
  Vlassis]{oliehoek2008optimal}
Frans~A Oliehoek, Matthijs~TJ Spaan, and Nikos Vlassis.
\newblock Optimal and approximate q-value functions for decentralized {POMDP}s.
\newblock \emph{Journal of Artificial Intelligence Research}, 32:\penalty0
  289--353, 2008.

\bibitem[Packer et~al.(2018)Packer, Gao, Kos, Kr{\"a}henb{\"u}hl, Koltun, and
  Song]{packer2018assessing}
Charles Packer, Katelyn Gao, Jernej Kos, Philipp Kr{\"a}henb{\"u}hl, Vladlen
  Koltun, and Dawn Song.
\newblock Assessing generalization in deep reinforcement learning.
\newblock \emph{arXiv preprint arXiv:1810.12282}, 2018.

\bibitem[Papoudakis et~al.(2021)Papoudakis, Christianos, Sch{\"a}fer, and
  Albrecht]{papoudakis2021benchmarking}
Georgios Papoudakis, Filippos Christianos, Lukas Sch{\"a}fer, and Stefano~V
  Albrecht.
\newblock Benchmarking multi-agent deep reinforcement learning algorithms in
  cooperative tasks.
\newblock In \emph{Thirty-fifth Conference on Neural Information Processing
  Systems Datasets and Benchmarks Track (Round 1)}, 2021.

\bibitem[Peng et~al.(2021)Peng, Li, Hui, Liu, and Zhou]{peng2021learning}
Zhenghao Peng, Quanyi Li, Ka~Ming Hui, Chunxiao Liu, and Bolei Zhou.
\newblock Learning to simulate self-driven particles system with coordinated
  policy optimization.
\newblock \emph{Advances in Neural Information Processing Systems},
  34:\penalty0 10784--10797, 2021.

\bibitem[Perolat et~al.(2022)Perolat, de~Vylder, Hennes, Tarassov, Strub,
  de~Boer, Muller, Connor, Burch, Anthony, et~al.]{perolat2022mastering}
Julien Perolat, Bart de~Vylder, Daniel Hennes, Eugene Tarassov, Florian Strub,
  Vincent de~Boer, Paul Muller, Jerome~T Connor, Neil Burch, Thomas Anthony,
  et~al.
\newblock Mastering the game of stratego with model-free multiagent
  reinforcement learning.
\newblock \emph{arXiv preprint arXiv:2206.15378}, 2022.

\bibitem[Qiu et~al.(2022)Qiu, Wang, Wang, An, Hu, Obraztsova, Rabinovich, Hao,
  Chen, and Fan]{qiu2022off}
Wei Qiu, Weixun Wang, Rundong Wang, Bo~An, Yujing Hu, Svetlana Obraztsova,
  Zinovi Rabinovich, Jianye Hao, Yingfeng Chen, and Changjie Fan.
\newblock Off-beat multi-agent reinforcement learning.
\newblock \emph{arXiv preprint arXiv:2205.13718}, 2022.

\bibitem[Rashid et~al.(2018)Rashid, Samvelyan, Schroeder, Farquhar, Foerster,
  and Whiteson]{rashid2018qmix}
Tabish Rashid, Mikayel Samvelyan, Christian Schroeder, Gregory Farquhar, Jakob
  Foerster, and Shimon Whiteson.
\newblock {QMIX}: Monotonic value function factorisation for deep multi-agent
  reinforcement learning.
\newblock In \emph{International Conference on Machine Learning}, pp.\
  4295--4304, 2018.

\bibitem[Rashid et~al.(2020)Rashid, Farquhar, Peng, and
  Whiteson]{rashid2020weighted}
Tabish Rashid, Gregory Farquhar, Bei Peng, and Shimon Whiteson.
\newblock Weighted qmix: Expanding monotonic value function factorisation for
  deep multi-agent reinforcement learning.
\newblock \emph{Advances in Neural Information Processing Systems},
  33:\penalty0 10199--10210, 2020.

\bibitem[Schulman et~al.(2015)Schulman, Moritz, Levine, Jordan, and
  Abbeel]{schulman2015high}
John Schulman, Philipp Moritz, Sergey Levine, Michael Jordan, and Pieter
  Abbeel.
\newblock High-dimensional continuous control using generalized advantage
  estimation.
\newblock \emph{arXiv preprint arXiv:1506.02438}, 2015.

\bibitem[Schulman et~al.(2017)Schulman, Wolski, Dhariwal, Radford, and
  Klimov]{schulman2017proximal}
John Schulman, Filip Wolski, Prafulla Dhariwal, Alec Radford, and Oleg Klimov.
\newblock Proximal policy optimization algorithms.
\newblock \emph{arXiv preprint arXiv:1707.06347}, 2017.

\bibitem[Silver et~al.(2018)Silver, Hubert, Schrittwieser, Antonoglou, Lai,
  Guez, Lanctot, Sifre, Kumaran, Graepel, et~al.]{silver2018general}
David Silver, Thomas Hubert, Julian Schrittwieser, Ioannis Antonoglou, Matthew
  Lai, Arthur Guez, Marc Lanctot, Laurent Sifre, Dharshan Kumaran, Thore
  Graepel, et~al.
\newblock A general reinforcement learning algorithm that masters chess, shogi,
  and go through self-play.
\newblock \emph{Science}, 362\penalty0 (6419):\penalty0 1140--1144, 2018.

\bibitem[Song et~al.(2019)Song, Jiang, Tu, Du, and
  Neyshabur]{song2019observational}
Xingyou Song, Yiding Jiang, Stephen Tu, Yilun Du, and Behnam Neyshabur.
\newblock Observational overfitting in reinforcement learning.
\newblock In \emph{International Conference on Learning Representations}, 2019.

\bibitem[Stone \& Kraus(2010)Stone and Kraus]{stone2010teach}
Peter Stone and Sarit Kraus.
\newblock To teach or not to teach?: decision making under uncertainty in ad
  hoc teams.
\newblock In \emph{AAMAS}, pp.\  117--124, 2010.

\bibitem[Stooke et~al.(2021)Stooke, Mahajan, Barros, Deck, Bauer, Sygnowski,
  Trebacz, Jaderberg, Mathieu, et~al.]{team2021open}
Adam Stooke, Anuj Mahajan, Catarina Barros, Charlie Deck, Jakob Bauer, Jakub
  Sygnowski, Maja Trebacz, Max Jaderberg, Michael Mathieu, et~al.
\newblock Open-ended learning leads to generally capable agents.
\newblock \emph{arXiv preprint arXiv:2107.12808}, 2021.

\bibitem[Sugden(2005)]{sugden2005rights}
Robert Sugden.
\newblock Rights, co-operation and welfare.
\newblock In \emph{The Economics of Rights, Co-operation and Welfare}, pp.\
  170--182. Springer, 2005.

\bibitem[Sutton \& Barto(2018)Sutton and Barto]{sutton2018reinforcement}
Richard~S Sutton and Andrew~G Barto.
\newblock \emph{Reinforcement Learning: An Introduction}.
\newblock MIT press, 2018.

\bibitem[Sutton et~al.(1999)Sutton, Precup, and Singh]{sutton1999between}
Richard~S Sutton, Doina Precup, and Satinder Singh.
\newblock Between mdps and semi-mdps: A framework for temporal abstraction in
  reinforcement learning.
\newblock \emph{Artificial intelligence}, 112\penalty0 (1-2):\penalty0
  181--211, 1999.

\bibitem[Sutton(1984)]{sutton1984temporal}
Richard~Stuart Sutton.
\newblock \emph{Temporal credit assignment in reinforcement learning}.
\newblock PhD thesis, University of Massachusetts Amherst, 1984.

\bibitem[Tampuu et~al.(2017)Tampuu, Matiisen, Kodelja, Kuzovkin, Korjus, Aru,
  Aru, and Vicente]{tampuu2017multiagent}
Ardi Tampuu, Tambet Matiisen, Dorian Kodelja, Ilya Kuzovkin, Kristjan Korjus,
  Juhan Aru, Jaan Aru, and Raul Vicente.
\newblock Multiagent cooperation and competition with deep reinforcement
  learning.
\newblock \emph{PLoS ONE}, 12\penalty0 (4), 2017.

\bibitem[Vezhnevets et~al.(2020)Vezhnevets, Wu, Eckstein, Leblond, and
  Leibo]{vezhnevets2020options}
Alexander Vezhnevets, Yuhuai Wu, Maria Eckstein, R{\'e}mi Leblond, and Joel~Z
  Leibo.
\newblock Options as responses: Grounding behavioural hierarchies in
  multi-agent reinforcement learning.
\newblock In \emph{International Conference on Machine Learning}, pp.\
  9733--9742. PMLR, 2020.

\bibitem[Vinyals et~al.(2019)Vinyals, Babuschkin, Czarnecki, Mathieu, Dudzik,
  Chung, Choi, Powell, Ewalds, Georgiev, et~al.]{vinyals2019grandmaster}
Oriol Vinyals, Igor Babuschkin, Wojciech~M Czarnecki, Micha{\"e}l Mathieu,
  Andrew Dudzik, Junyoung Chung, David~H Choi, Richard Powell, Timo Ewalds,
  Petko Georgiev, et~al.
\newblock Grandmaster level in {StarCraft II} using multi-agent reinforcement
  learning.
\newblock \emph{Nature}, 575\penalty0 (7782):\penalty0 350--354, 2019.

\bibitem[Wang et~al.(2021{\natexlab{a}})Wang, Ren, Liu, Yu, and
  Zhang]{wang2021qplex}
Jianhao Wang, Zhizhou Ren, Terry Liu, Yang Yu, and Chongjie Zhang.
\newblock {QPLEX}: Duplex dueling multi-agent q-learning.
\newblock In \emph{International Conference on Learning Representations},
  2021{\natexlab{a}}.

\bibitem[Wang et~al.(2018)Wang, Hao, Wang, and Taylor]{wang2018towards}
Weixun Wang, Jianye Hao, Yixi Wang, and Matthew Taylor.
\newblock Towards cooperation in sequential prisoner's dilemmas: a deep
  multiagent reinforcement learning approach.
\newblock \emph{arXiv preprint arXiv:1803.00162}, 2018.

\bibitem[Wang et~al.(2021{\natexlab{b}})Wang, Han, Wang, Dong, and
  Zhang]{wang2021dop}
Yihan Wang, Beining Han, Tonghan Wang, Heng Dong, and Chongjie Zhang.
\newblock {DOP}: Off-policy multi-agent decomposed policy gradients.
\newblock In \emph{International Conference on Learning Representations},
  2021{\natexlab{b}}.

\bibitem[Watkins \& Dayan(1992)Watkins and Dayan]{watkins1992q}
Christopher~JCH Watkins and Peter Dayan.
\newblock Q-{L}earning.
\newblock \emph{Machine Learning}, 8\penalty0 (3-4):\penalty0 279--292, 1992.

\bibitem[Williams(1992)]{williams1992simple}
Ronald~J Williams.
\newblock Simple statistical gradient-following algorithms for connectionist
  reinforcement learning.
\newblock \emph{Machine learning}, 8\penalty0 (3):\penalty0 229--256, 1992.

\bibitem[Yang \& Wang(2020)Yang and Wang]{yang2020overview}
Yaodong Yang and Jun Wang.
\newblock An overview of multi-agent reinforcement learning from game
  theoretical perspective.
\newblock \emph{arXiv preprint arXiv:2011.00583}, 2020.

\bibitem[Yu et~al.(2021)Yu, Velu, Vinitsky, Wang, Bayen, and
  Wu]{yu2021surprising}
Chao Yu, Akash Velu, Eugene Vinitsky, Yu~Wang, Alexandre Bayen, and Yi~Wu.
\newblock The surprising effectiveness of ppo in cooperative, multi-agent
  games.
\newblock \emph{arXiv preprint arXiv:2103.01955}, 2021.

\bibitem[Yuan et~al.(2022)Yuan, Wang, Zhang, Wang, Zhang, Yu, and
  Zhang]{yuan2022multi}
Lei Yuan, Jianhao Wang, Fuxiang Zhang, Chenghe Wang, ZongZhang Zhang, Yang Yu,
  and Chongjie Zhang.
\newblock Multi-agent incentive communication via decentralized teammate
  modeling.
\newblock \emph{Proceedings of the AAAI Conference on Artificial Intelligence},
  36\penalty0 (9):\penalty0 9466--9474, Jun. 2022.

\bibitem[Zha et~al.(2020)Zha, Ma, Yuan, Hu, and Liu]{zha2020rank}
Daochen Zha, Wenye Ma, Lei Yuan, Xia Hu, and Ji~Liu.
\newblock Rank the episodes: A simple approach for exploration in
  procedurally-generated environments.
\newblock In \emph{International Conference on Learning Representations}, 2020.

\bibitem[Zhang et~al.(2021)Zhang, Yang, and Ba{\c{s}}ar]{zhang2021multi}
Kaiqing Zhang, Zhuoran Yang, and Tamer Ba{\c{s}}ar.
\newblock Multi-agent reinforcement learning: A selective overview of theories
  and algorithms.
\newblock \emph{Handbook of Reinforcement Learning and Control}, pp.\
  321--384, 2021.

\end{thebibliography}
\bibliographystyle{iclr2023_conference}

\clearpage

\appendix

\section{Environments}\label{secapp:environments}

This section introduces Melting Pot in-depth, including substrates and evaluation scenarios.

\subsection{General Settings}

\begin{wrapfigure}{r}{0.42\textwidth}
    \begin{center}
        \includegraphics[width=.42\textwidth]{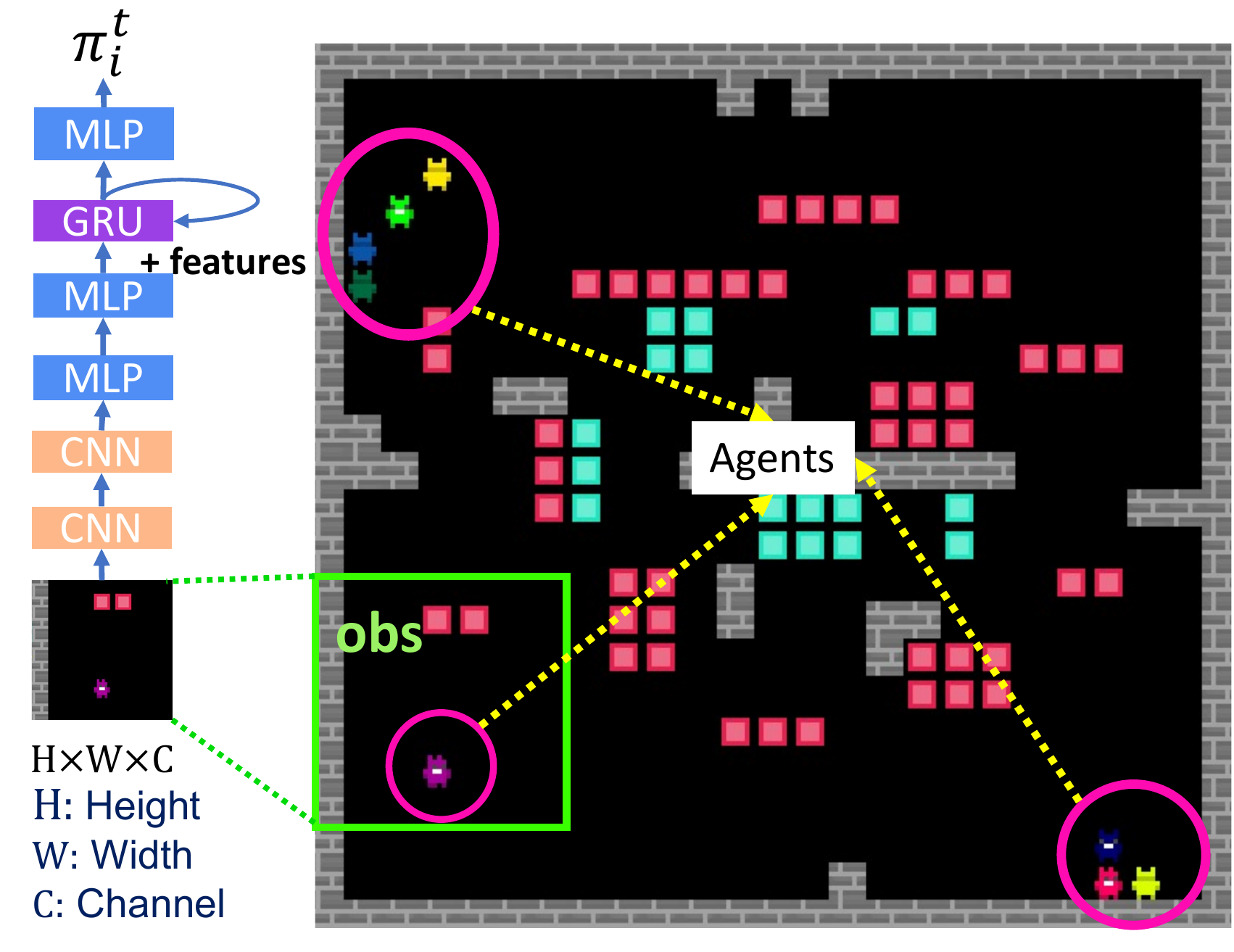}
        \caption{The green box to the lower left shows the agent's observation.}\label{fig:mp_example_appendix}
    \end{center}
\end{wrapfigure}
Melting Pot~\citep{leibo2021scalable} is a suite of testbeds for MARL evaluation. It proposes a novel evaluation pipeline for evaluating the MARL method in various domains. That is, all MARL agents are trained in the substrate; during evaluation, some agents are selected as the focal agents (the agents to be evaluated), and the rest agents become the background agents (pretrained policies of MARL models will be plugged in); the evaluation scenarios share the same physical properties with the substrates. Melting Pot environments possess many properties, such as temporal coordination and free riding. MARL agent performing well in these environments means its behaviors demonstrate these properties. In each substrate, episodes last 1000 or 2000 steps. The agents have a partial observability window of $11\times11$ sprites. The agent can observe 9 rows in front of itself, 1 row behind, and 5 columns to either side. Sprites are $8\times8$ pixels. Thus, in RGB pixels, the size of each observation is $88\times88\times3$. All agents use RGB pixel representations as their inputs. In ~\autoref{fig:mp_example_appendix}, the agent's observation is shown in the green box to the lower left of the state (\ie, the whole image). The agent is in the lower middle of the observation. The neural network architecture of the agent's policy is shown on the left. We introduce the neural network architecture design in Appendix~\ref{secapp:architectures}, MARL training and hyperparameters in Appendix~\ref{secapp:marl_training}.

\begin{figure}[ht]
    \centering
    \includegraphics[scale=0.3855]{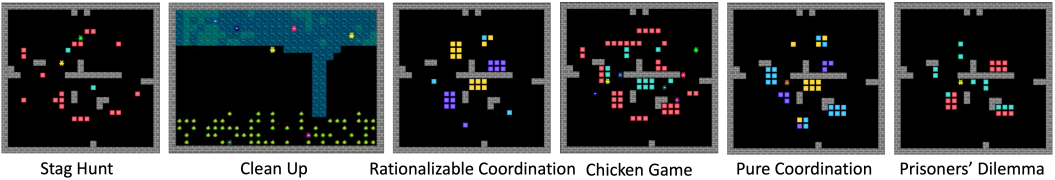}
    \caption{Melting Pot environments.}
    \label{fig:mp_scenarios_appdenxi}
\end{figure}

\subsection{Substrates and Evaluation Scenarios}
We introduce substrates and evaluation scenarios used in the experiments. In all substrates and scenarios, agents' movement actions are: \texttt{forward}, \texttt{backward}, \texttt{strafe left}, \texttt{strafe right}, \texttt{turn left}, \texttt{turn right}. Unless otherwise stated, each episode lasts 1000 steps. We show the environments in Figure~\ref{fig:mp_scenarios_appdenxi}, for readers' convenience.

\textbf{Chicken Game.} In this environment, there are 8 agents in the substrate. Agents move around the environments\footnote{There are two categories of environments: substrates and evaluation scenarios.} and collect resources of 2 different colors. Each agent carries an inventory $\rho = (\rho_1, \rho_2)$ with the count of resources picked up since the last respawn. Due to partial observability, agents can only observe their inventory\footnote{It also applies to other environments where agents have inventories}. The more resources of a given type an agent picks up, the more committed the agent becomes to the pure strategy corresponding to that resource\footnote{It also applies for other environments where matrix games should be resolved when two-agent interactions occur.}. The agent can zap the other agent via its zapping beam for interaction. When an interaction occurs, a traditional matrix game is started. Here, in this environment, it is a Chicken Game~\citep{sugden2005rights} where both agents trying to exploit the other leads to the worst payoff, \ie rewards, for both. Gathering red resources makes the agent's strategy towards committing `hawk' while collecting green resources pushes it toward playing `dove'. The payoff matrix for row and column players is:
\begin{equation}
    \Phi_{\text{row}} = \Phi^{\text{T}}_{\text{col}} = 
    \begin{bmatrix}
    3 & 2 \\
    5 & 0
    \end{bmatrix}.
\nonumber
\end{equation}

\textbf{Chicken Game (eval).} The task and the payoff matrix in this scenario are the same as in Chicken Game. In this scenario, one focal agent is joining seven background agents. Unlike the focal agent that can play any strategy, the background agents were pretrained with pseudo rewards to play `dove'. The best strategy for the focal agent is to play `hawk'.

\textbf{Stag Hunt.} Similar to Chicken Game, there are 8 agents in this environment. Each agent collects resources that represent `hare' (red) or `stag' (green) and compares inventories in an interaction, \ie, encounter. The results of solving the encounter are the same as the classic Stag Hunt matrix game. In this environment, agents are facing tension between the reward for the team and the risk for the individual. The matrix for the interaction is:
\begin{equation}
    \Phi_{\text{row}} = \Phi^{\text{T}}_{\text{col}} = 
    \begin{bmatrix}
    4 & 0 \\
    2 & 2
    \end{bmatrix}.
\nonumber
\end{equation}

\textbf{Stag Hunt (eval).} In this environment, one agent interacts with seven pretrained agents. All background agents were trained to play the `stag' strategy during the interaction. The optimal policy for the focal agent is also to play `stag'.

\textbf{Clean Up.} There are seven agents in the environment. Agents are rewarded (+1) for collecting apples. In the environment, there are an orchard and a river. Agents should clean the river frequently to reduce pollution for the irrigation of the orchard. Apples in the orchard grow at a rate inversely related to the river's cleanliness. When the cleanliness rate reaches a certain threshold, apples stop growing. Agents can take \texttt{clean} action to clean a small amount of pollution from the river. However, such action only works in a small region around the agent in the river. So, agents should move to clean the river without any rewards. Consequently, agents should maintain the public good of orchard regrowth by cleaning the river. This creates a tension between the short-term individual incentive to maximize agents' reward by staying in the orchard and the long-term group interest in a clean river.

\textbf{Clean Up (eval).} In this evaluation scenario, three focal agents join four background agents. All background agents have been trained to behave altruistically, \ie, always cleaning the river without consuming apples. Thus, the optimal policy for the focal agent is to collect as many apples as possible without moving out of the orchard to clean the river.

\textbf{Pure Coordination.} In this environment, eight agents cannot be identified as individuals because all agents look the same. Agents gather resources of three different colors. So, the size of the agent's inventory is 3. To maximize the reward, all agents should collect the same colored resource when the encounter occurs. The matrix for the interaction is:
\begin{equation}
    \Phi_{\text{row}} = \Phi^{\text{T}}_{\text{col}} = 
    \begin{bmatrix}
    1 & 0 & 0 \\
    0 & 1 & 0 \\
    0 & 0 & 1
    \end{bmatrix}.
\nonumber
\end{equation}

\textbf{Pure Coordination (eval).} In this evaluation scenario, there are seven focal agents and one background agent. The background agent has been trained to target one particular resource out of three colors of resources. Focal agents should observe other agents to see the resources other agents are collecting and then decide the right color to pick. This scenario aims to evaluate that agents' coordination is not disrupted by the presence of unfamiliar other agents who has a special preference for one particular colored resource.

\textbf{Prisoners' Dilemma.} Eight agents collect colored resources that represent `defect' (red) or `cooperate' (green). Agents compare their inventories in an encounter where a classic Prisoner's Dilemma matrix game is resolved. Agents face tension between the reward for the group and the reward for the individual. The matrix for the interaction is:
\begin{equation}
    \Phi_{\text{row}} = \Phi^{\text{T}}_{\text{col}} = 
    \begin{bmatrix}
    3 & 0 \\
    4 & 1
    \end{bmatrix}.
\nonumber
\end{equation}

\textbf{Prisoners' Dilemma (eval).} In this evaluation scenario, one focal agent joins seven background agents. All background agents will play cooperative strategies, \ie, collecting `cooperate' resources and rarely collecting `defect').  The optimal policy for the focal agent is to identify such a pattern and then collect `defect' resources.

\textbf{Rational Coordination.} The environment setting is the same as Pure Coordination, except that different colored resources are of different values. Agents should find the optimal color to maximize the group reward. The matrix for the interaction is:
\begin{equation}
    \Phi_{\text{row}} = \Phi^{\text{T}}_{\text{col}} = 
    \begin{bmatrix}
    1 & 0 & 0\\
    0 & 2 & 0\\
    0 & 2 & 3
    \end{bmatrix}.
\nonumber
\end{equation}

\textbf{Rational Coordination (eval).} In this evaluation scenario, there are seven focal agents and one background agent. The background agent has been trained to target one particular resource out of three colors of resources. This scenario is similar to \textbf{Pure Coordination (eval)} since it aims to evaluate that agents’ coordination is not disrupted by the presence of unfamiliar other agents who has a special preference for one particular colored resource. However, this scenario is more challenging than \textbf{Pure Coordination (eval)}. While focal agents' choices are better than miscoordination, some choices are better than coordinating for the focal agents.

\newpage
\section{Baselines}\label{secapp:baselines}

We introduce baselines trained and evaluated in the experiment in detail. Baselines are MAPPO~\citep{yu2021surprising}, MAA2C~\citep{papoudakis2021benchmarking}, OPRE~\citep{vezhnevets2020options}, RandNet~\citep{lee2019network} and HFSP~\citep{heinrich2015fictitious,baker2019emergent}.

\subsection{MAPPO}

MAAPO is an extension of PPO~\citep{schulman2017proximal} for multi-agent RL. Following the CTDE~\citep{oliehoek2008optimal} training and execution paradigm, agents take actions independently during execution and agents' policies are trained via sharing information (\eg,) with other agents. In MAPPO, there are $N$ policies $\{ \pi_i \}^{N}_{i=1}$ for each agent $i$. A central critic is maintained by feeding all agents' observations and actions $\{o^{t}_{i}, u^{t}_{i}\}^{N}_{i=1}$. Although the global state contains all agents' observations, it contains redundant information that deteriorates the central critic learning with TD-learning~\citep{sutton1984temporal}. Note that all baselines that have a central critic takes all agents' observations and actions $\{o^{t}_{i}, u^{t}_{i}\}^{N}_{i=1}$ as the input.

\subsection{MAA2C}

MAA2C is a multi-agent RL variant of A2C~\citep{mnih2016asynchronous}. MAA2C adopts the same training and execution paradigm used in MAPPO. Similar to A2C, TD error is used  as the advantage in MAA2C for training agents' policies via maximizing policy gradient loss.

\subsection{OPRE}

We build OPRE~\citep{vezhnevets2020options} on top of MAPPO. The key idea behind OPRE is to re-use the same latent space to factorise the policy via creating a hierarchical policy structure:
\begin{equation}
    \pi_{i} (u_{i} | o^{\leq t}_{i}, o^{\prime}) = \sum_{z} q (z|o^{\prime}) \eta (u_i | o^{\leq t}_{i}, z) \nonumber
\end{equation}
where $o^{\prime}$ is $\{o^{t}_{i}, u^{t}_{i}\}^{N}_{i=1}$ and $\eta (u_i | o^{\leq t}_{i}, z)$ is a mixture component of the policy, \ie, an option. $o^{\leq t}_{i}$ can be represented via recurrent neural networks~\citep{hochreiter1997long,cho2014properties}.  Note that the `option' here differs from the \textit{option} in hierarchical RL~\citep{sutton1999between,bacon2017option}. In OPRE, the option has no explicit probability distribution of entering an option and no explicit probability distribution of exiting the current option. The behavior policy is defined as:
\begin{equation}
    \mu_{i} (o^{\leq t}_{i}) = \sum_{z} p (z|o^{\prime}) \eta (u_i | o^{\leq t}_{i}, z) \nonumber
\end{equation}
Then $p (z|o^{\prime})$ can be trained via $\texttt{KL}(q||p)$ together with the policy and the central critic in an end-to-end manner. We use the default hyperparameters used in OPRE in our experiments. The number of options is $16$.

\subsection{RandNet}

\cite{lee2019network} proposed RandNet for improving the generalization of RL in unseen environments, especially environments with new textures and layouts. RandNet utilizes a single-layer convolutional neural network (CNN) as a random network,  where its output has the same dimension with the input. To reinitialize the parameters of the random network, RandNet utilizes the following mixture of distributions: $P(\phi) = \alpha \mathbb{I} (\phi=\mathbf{I}) + (1-\alpha)\mathcal{N}\left(\mathbf{0};\sqrt{\frac{2}{n_{\text{in}} + n_{\text{out}}}}\right)$ where $\mathbf{I}$ is an identity kernel, $\alpha \in [0, 1]$ is a positive constant, $\mathcal{N}$ stands for the normal distribution. $n_{\text{in}}$ and $n_{\text{out}}$ are the number of input and output channels, respectively. We use RandNet in the policy network and the critic network of MAPPO. We use the default hyperparameters used in RandNet in our experiments.

\subsection{HFSP}

Self-play~\citep{brown1951iterative,heinrich2015fictitious,silver2018general,baker2019emergent} has been studied for obtaining equilibria via creating fictitious plays by sampling agents' past policies. HFSP is a heuristic fictitious self-play method. HFSP uses the MARL framework of MAPPO. Like RPM, it maintains a memory to save all the policies after each training step. HFSP agents have a probability of $0.7$ to sample the lasted policies and a probability of $0.3$ to sample previous policies. RPM can be considered as a ranked self-play by sampling policies with a hierarchy. 

\newpage
\section{Architectures}\label{secapp:architectures}

We first introduce the neural network architecture of the policy, the critic and the training pipeline for all methods, and the hyperparameters used in the neural network architectures. RPM and all baselines use the same network architecture.

\begin{figure}[ht]
    \centering
    \includegraphics[scale=0.5]{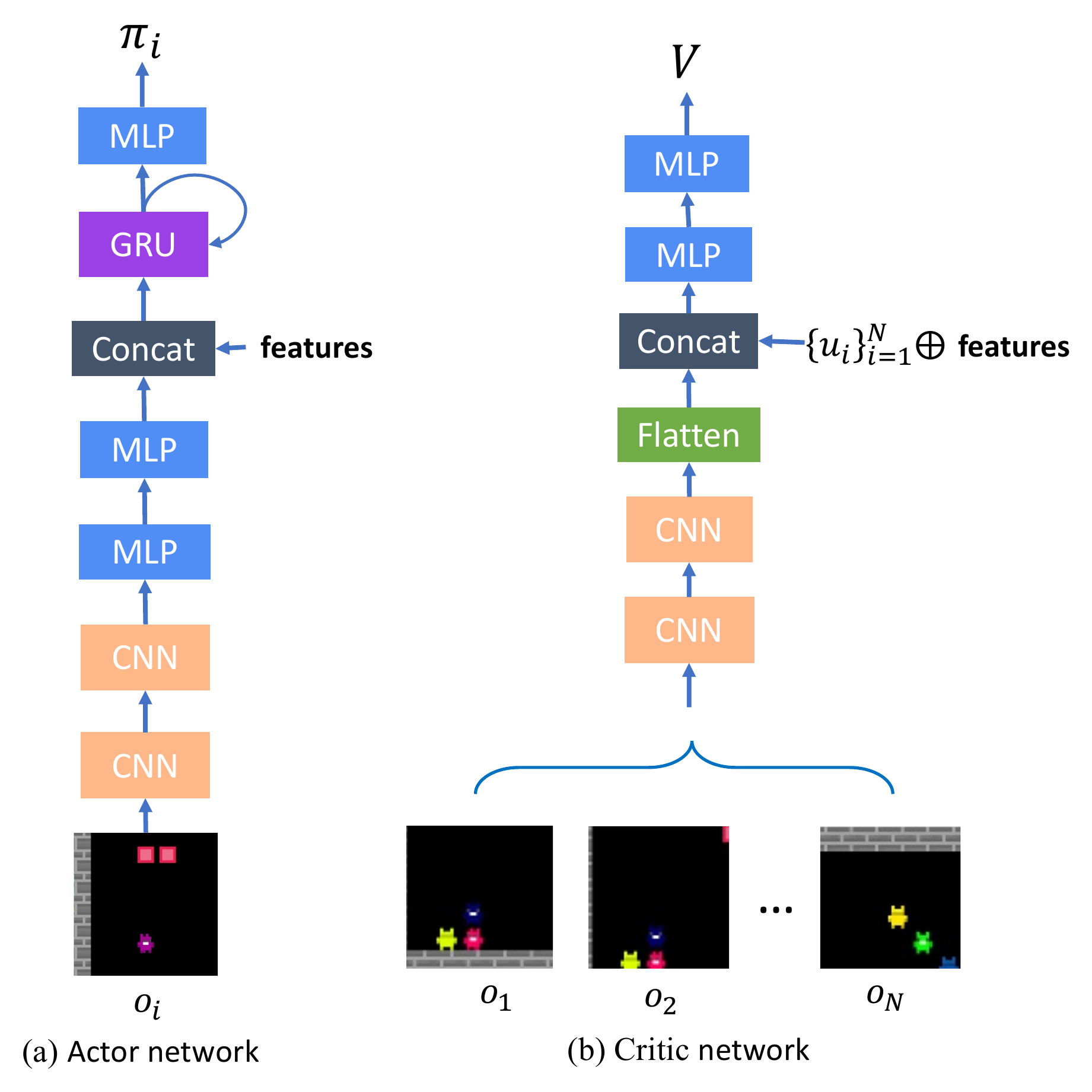}
    \caption{The networks of the policy (left) and the critic (right).}
    \label{fig:marl_network_appendix}
\end{figure}

\textbf{Actor Network.} The actor network consists of a convolutional neural network (CNN) with two layers. The two CNN layers use the ReLU activation function. The first and the second layer have 16 and 32 output channels, 8 and 4 kernel shapes and 8 and 1 strides, respectively. An MLP follows the two CNN layers with two layers with 64 neurons each. The MLP uses the ReLU action function. It is then followed by a GRU~\citep{cho2014properties} with 128 units. The input of the GRU is the concatenation of the output of the MLP and the features (such as the agent's position, orientation and inventory). The output of the GRU is fed into the MLP, and it outputs the policy $\pi_i$ for agent $i$.
 
\textbf{Critic Network.} The critic network is shared by all agents. The critic network consists of a CNN with two layers. The two CNN layers use the ReLU activation function. The first and the second layer have 16 and 32 output channels, 8 and 4 kernel shapes and 8 and 1 strides, respectively. The CNN is then followed by a concatenation of all agents' actions and features (such as agent's position, orientation and inventory). The concatenation is then fed into an MLP with two layers with 64 neurons. The MLP uses the ReLU action function. The MLP outputs the value, a vector with the dimension of $N$. We take all agents' observations as a batch and feed them into the CNN. We then flatten the CNN's output and feed it with agents' actions and features as inputs to the MLP network to get the value vector for all agents. 

\textbf{Training.} Our training framework is a distributed framework with 30 CPU cores to collect experiences and 1 GPU for the learner to learn policies, similar to the framework used in IMPALA~\citep{espeholt2018impala}. To improve the efficiency and save memory, we use parameter sharing~\citep{rashid2018qmix,wang2021qplex,yu2021surprising}, \ie, all agents share a policy network. We adopt the CTDE framework to train the policies and the critic.
 
\newpage
\section{Training Settings}\label{secapp:marl_training}

We implement our method with Python and PyTorch. The learner is implemented with  EPyMARL~\citep{papoudakis2021benchmarking} and the actors that collect experiments are implemented with Ray~\citep{moritz2018ray}. We train agents in Melting Pot substrates for 200 million frames with 3 random seeds for RPM and 4 seeds for baselines. We randomly sample policies from RPM. The discount factor $\gamma=0.99$ and we follow the default hyper-parameters used in the original papers of all methods in our research. We carry out experiments on NVIDIA A100 Tensor Core GPU. We resort to mean-std values as our performance evaluation measurement. We use Adam as our optimizer. We list some important hyper-parameters in Table.~\ref{tab:hyparam}.

\begin{table*}[ht]
\caption{Hyper-parameters}\label{tab:hyparam}
\centering
\begin{tabular}{{c}{c}}
    \toprule
    \textbf{hyper-parameter} & \textbf{Value} \\
    \midrule
    Optimizer & Adam \\
    Learning rate  & 1e-4 \\
    Adam betas & $(0.9, 0.999)$\\
    Adam epsilon & 1e-8 \\
    Adam weight decay & 0 \\
    Gradient norm clip & 10 \\
    \midrule
    Batch size & 60 \\
    Replay buffer size &  600 \\
    \midrule
    $k$ in $k$-step return  & 5 \\
    $\gamma$ & 0.99 \\
    Evaluation interval & 1,000 \\
    Target update interval & 200 \\
    \midrule
    $p$ & 0.5 \\
    \bottomrule
\end{tabular}
\end{table*}

\begin{table*}[ht]
\caption{The value of $\psi$}\label{tab:psi_table}
\centering
\begin{tabular}{{c}{c}}
    \toprule
    \textbf{Melting Pot Substrate} & \textbf{The value of} $\psi$ \\
    \midrule
    Stag Hunt & 1 \\
    Pure Coordination & 0.01 \\
    Clean Up & 1 \\
    Prisoners' Dilemma & 0.02 \\
    Rational Coordination & 0.2 \\
    Chicken Game & 1 \\
    \bottomrule
\end{tabular}
\end{table*}

\newpage

\section{Results}\label{secapp:results}

\begin{figure}[ht]
\begin{center}
\includegraphics[scale=0.375]{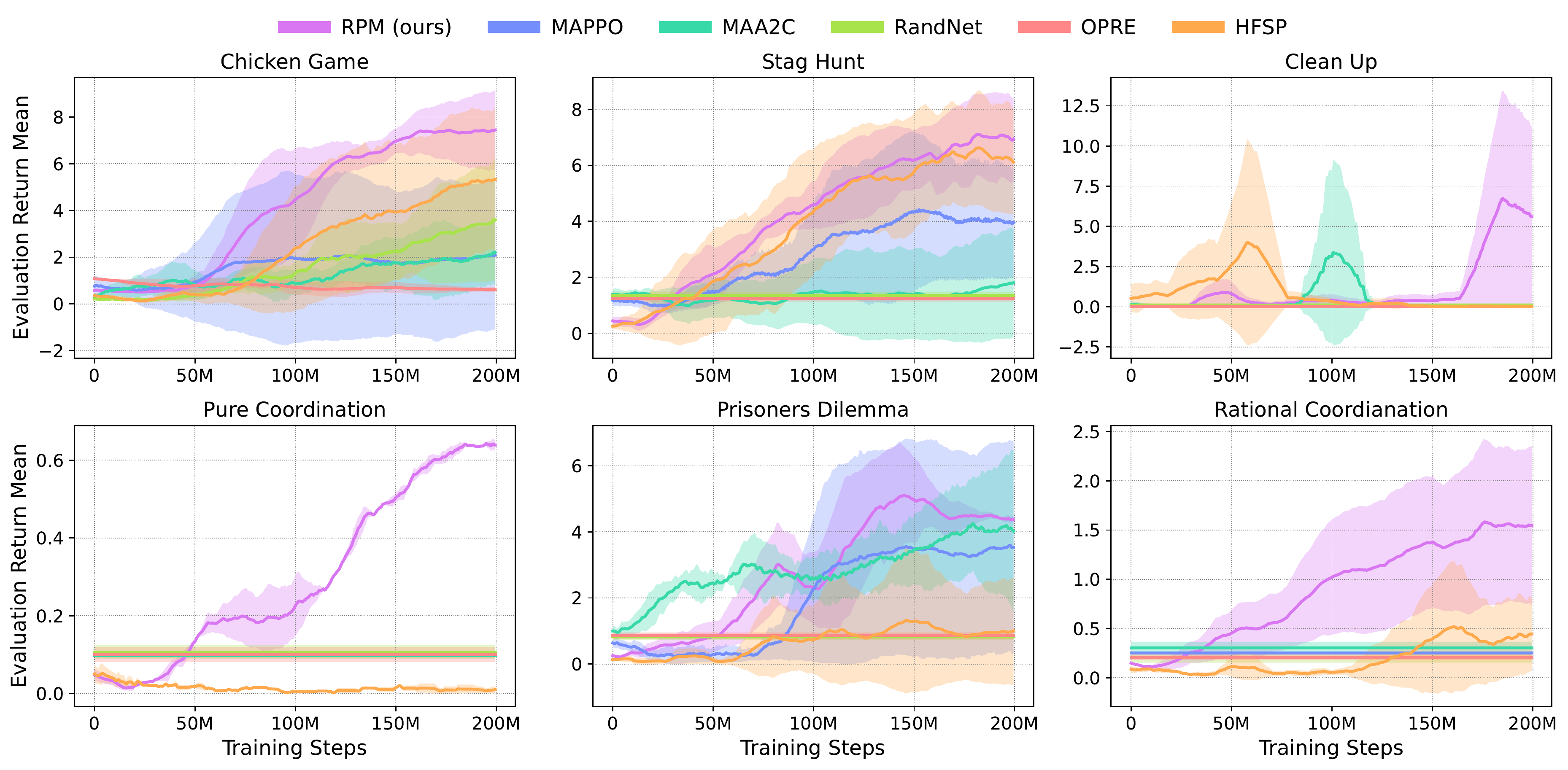}
\end{center}
\caption{Episode return in substrates.}
\label{fig:exp_mp_baselines_train}
\end{figure}

We depict the episode return within the substrate. During training, the MARL methods are evaluated in the substrate. Figure~\ref{fig:exp_mp_baselines_train} demonstrates that despite the environments being distinct, RPM also demonstrates leading performance. Once the agents in the substrate achieve a satisfactory episode return, the trained policy will be saved at the appropriate rank. In turn, it improves the performance of RPM in the evaluation scenario by collecting diverse data on multi-agent interactions.

\end{document}